\documentclass[%
 reprint,
 amsmath,amssymb,
 aps,
 pra,
]{revtex4-2}
\usepackage{changes}
\definecolor{addedColor}{rgb}{0,0.6,0} % 绿色
\definecolor{deletedColor}{rgb}{1,0,0} % 红色
\definecolor{highlightColor}{rgb}{1,1,0} % 黄色
\definecolor{commentColor}{rgb}{0,0,1} % 蓝色

\setaddedmarkup{\textcolor{addedColor}{#1}} % 添加的文本用绿色
\setdeletedmarkup{\textcolor{deletedColor}{\sout{#1}}} % 删除的文本用红色划掉
\sethighlightmarkup{\colorbox{highlightColor}{#1}} % 高亮文本黄色背景
\setcommentmarkup{\textcolor{commentColor}{[#1]}} % 评论用蓝色

\usepackage{float}
\usepackage{tikz}
\usepackage{lipsum}
\usepackage{braket}
\usepackage{tabularx}
\usepackage{float}
\usepackage{subfigure}

\usepackage{graphicx}% Include figure files
\usepackage{dcolumn}% Align table columns on decimal point
\usepackage{bm}% bold math

\usepackage{braket}
\usepackage{xcolor}
\usepackage{graphicx}

\begin{document}

\preprint{APS/123-QED}

\title{Molecular docking via quantum approximate optimization algorithm}
%\title{Quantum-Assisted Molecular Docking: A Novel Approach Using Quantum Approximate Optimization Algorithm}
% Force line breaks with \\

\author{Qi-Ming Ding}
\email{dqiming94@pku.edu.cn}
\affiliation{Center on Frontiers of Computing Studies, Peking University, Beĳing 100871, China}%
\affiliation{School of Computer Science, Peking University, Beĳing 100871, China}%

\author{Yi-Ming Huang }
\email{ymhwang@pku.edu.cn}
\affiliation{Center on Frontiers of Computing Studies, Peking University, Beĳing 100871, China}%
\affiliation{School of Computer Science, Peking University, Beĳing 100871, China}%

\author{Xiao Yuan}
\email{xiaoyuan@pku.edu.cn}
\affiliation{Center on Frontiers of Computing Studies, Peking University, Beĳing 100871, China}%
\affiliation{School of Computer Science, Peking University, Beĳing 100871, China}%
\date{\today}% It is always \today, today,

\begin{abstract}
% to molecular docking by leveraging the Quantum Approximate Optimization Algorithm (QAOA) and its variant, the Digitized-counterdiabatic QAOA (DC-QAOA). [XXX add the examples considered in the main text.] \replaced{furnishing insights into protein functionalities and fostering the development of novel therapeutics.}{enabling us to understand protein functions and advance novel therapeutics.}\replaced{The DC-QAOA has demonstrated exceptional efficacy, yielding docking results of heightened accuracy and biological pertinence, particularly in complex molecular docking challenges. Additionally, algorithms founded on QAOA have shown increased compatibility with hardware in the context of the noisy intermediate-scale quantum (NISQ) era, suggesting their effectiveness in practical docking applications. }{The DC-QAOA exhibits superior performance, providing more accurate and biologically relevant docking results, especially for larger molecular docking problems. Moreover, QAOA-based algorithms demonstrate enhanced hardware compatibility in the noisy intermediate-scale quantum era, indicating their potential for efficient implementation under practical docking scenarios. } \replaced{Our research thus highlights the promising role of quantum computing in drug discovery and contributes significant insights towards the optimization of protein-ligand docking methodologies.}{Our findings underscore quantum computing's potential in drug discovery and offer valuable insights for optimizing protein-ligand docking processes.}

Molecular docking plays a pivotal role in drug discovery and precision medicine, furnishing insights into protein functionalities and fostering the development of novel therapeutics. Here, we introduce a potential alternative solution to this problem using QAOA-based algorithm. Our method was applied to analyze diverse biological systems, including the SARS-CoV-2 Mpro complex with PM-2-020B, the DPP-4 complex with piperidine fused imidazopyridine 34, and the HIV-1 gp120 complex with JP-III-048. We found that the digitized-counterdiabatic quantum approximate optimization algorithm (DC-QAOA), , which integrates the concept of counterdiabatic driving, surpasses the conventional QAOA in terms of quantum circuit depth and optimization efficiency. This is particularly evident in complex molecular docking challenges, where DC-QAOA delivers more accurate and biologically relevant results. Our research highlights the promising role of quantum computing in drug discovery and contributes significant insights towards the optimization of protein-ligand docking methodologies.

\end{abstract}
\maketitle

%\tableofcontents  the digitized-counterdiabatic quantum approximate optimization algorithm (DC-QAOA), which utilizes counterdiabatic driving and QAOA on a quantum computer. Additionally, algorithms founded on QAOA have shown increased compatibility with hardware in the context of the noisy intermediate-scale quantum (NISQ) era, suggesting their effectiveness in practical docking applications. 

\section{\label{sec:level1} Introduction}
%Unraveling the folding dynamics could yield potential treatments for neurodegenerative disorders, such as Alzheimer's, Huntington's, and Parkinson's, which arise from protein misfolding. 

Quantum computing, a concept first introduced in the 1980s by Feynman and other pioneers~\cite{feynman1982simulating,manin1980computable,benioff1980computer}, has made significant strides, particularly with the developments of Shor's algorithm~\cite{shor1994algorithms} and others~\cite{nielsen2002quantum}. Despite these advances, constructing a universal fault-tolerant quantum computer remains experimentally challenging. Currently, we are in the noisy intermediate-scale quantum (NISQ) era, where state-of-the-art quantum devices have a limited number of qubits and imperfect operations~\cite{preskill2018quantum}. Since the NISQ hardware is  weaker than a universal quantum computer~{~\cite{preskill2018quantum,barak2021classical,bharti2022noisy}}, it is widely believed that NISQ hardware can only be used to solve specific problems. Therefore, the identification of tasks that could leverage quantum features of NISQ devices is of great importance~\cite{bharti2022noisy,cheng2023noisy,kim_evidence_2023}.

One promising application of quantum computing lies in the realm of drug discovery~\cite{cao2018potential,zinner2021quantum}, encompassing vital areas such as protein folding and molecular docking. Protein folding, commonly approached through 2D or 3D lattice models to minimize interaction energy, can be elegantly reframed as a Hamiltonian problem. {By employing quantum annealing~\cite{perdomo2008construction,perdomo2012finding,irback2022folding} or gate-based quantum computation~\cite{babej2018coarse,fingerhuth2018quantum,robert2021resource,chandarana2023digitized}, we can reformulate protein folding as a ground state problem.
% the ground state can efficiently capture the protein configuration. ,saito2023lattice
Significant progress has been made in this field, with explorations of system sizes up to 9 amino acids utilizing various quantum hardware platforms~\cite{chandarana2023digitized}.}
% \yx{[XXX comment on how helpful quantum computing is in protein folding].}
Molecular docking, another potential field in quantum applications, can be broadly classified into score function-based approaches and score function-independent methods. Score function-based approaches involve scoring functions to evaluate docking poses, as demonstrated in recent studies~\cite{malone2022towards, kirsopp2022quantum}. On the other hand, score function-independent methods explore alternative strategies. For example, Gaussian Boson sampling enables the identification of docking postures based on the geometric relationship between proteins and ligands~\cite{banchi2020molecular,yu2022universal}. Another approach involves solving the molecular unfolding problem using quantum annealing, which serves as the first step in geometric molecular docking techniques~\cite{mato2022quantum}.

% \yx{[XXX Any limitations of the other works, such as score function-based approaches and maximizing internal atomic distances for molecular unfolding]}\replaced{ansatz}{Ansatz}
However, approaches based on scoring functions require precise computation of the binding energy between proteins and ligands. Even a slight error of 6 kJ/mol can cause significant deviations in the docking results~\cite{cao2018potential}, and achieving such accuracy with current quantum devices is difficult. Meanwhile, the other approach employing Gaussian Boson sampling does not provide a definitive guarantee of clique formation, and additional calculations are generally necessary to confirm its occurrence~\cite{banchi2020molecular}. Additionally, in practical docking scenarios, the adjacency matrix often exhibits sparsity, which can lead to experimental parameters approaching zero. As a result, achieving these parameters poses a challenge in optical experiments~\cite{yu2022universal}.

A potential solution to these challenges is to exploit digital quantum computing to solve the maximum weighted clique of a given graph, thereby addressing the molecular docking problem. Given the constraints of current quantum hardware, many NISQ algorithms adopt a hybrid quantum-classical strategy, known as variational quantum algorithms (VQAs)~\cite{cerezo2021variational}. The quantum approximate optimization algorithm (QAOA) is a prominent type of VQAs~\cite{farhi2014quantum}, which use shallow quantum circuits and classical optimization to find the optimal state corresponding to a $p$-level parametrized quantum circuit. As $p \to \infty$, the final state approaches an adiabatically evolved state and becomes an exact maximal eigenstate. However, increasing $p$ also leads to escalating classical optimization costs. Efforts to enhance QAOA primarily focus on reducing the quantum circuit depth~\cite{blekos2023review}, with improved versions such as ADAPT-QAOA~\cite{zhu2022adaptive}, digital-analog QAOA~\cite{Headley2022}, and digitized-counterdiabatic QAOA(DC-QAOA)~\cite{chandarana2022digitized}. In particular, DC-QAOA, incorporates a counterdiabatic (CD) driving term to develop a superior  {ansatz}, thereby enhancing global performance. The QAOA has been validated in various applications, including Ising spin models, entangled state preparation, classical optimization problems such as MaxCut and the SK model, and the P-spin model~\cite{farhi2014quantum,xiaoming2022}.

In this paper, we propose using digital quantum computing to tackle the molecular docking problem, thereby circumventing the limitations inherent to the Gaussian Boson sampling method. Drawing inspiration from previous literature~\cite{kuhl1984combinatorial, banchi2020molecular,yu2022universal}, we improved the criterion of constructing a protein-ligand binding interaction graph between specified proteins and ligands. Our objective is to identify the maximum vertex weight clique within this graph, which signifies the most probable docking posture. This approach uniquely translates a complex biological problem into a mathematical optimization problem that can be effectively addressed using digital quantum computers. We derive the cost Hamiltonian from mathematical optimization formulations of the maximum vertex weight clique and solve it using QAOA and DC-QAOA. From the ground state, we can then extract the optimal binding conformation. Our method introduces a novel paradigm for applying near-term digital quantum computers to the field of biopharmaceutics, underscoring the potential of quantum computing in solving intricate real-world problems.

\begin{figure*}[t]
    \centering
    \includegraphics[width=0.95\textwidth]{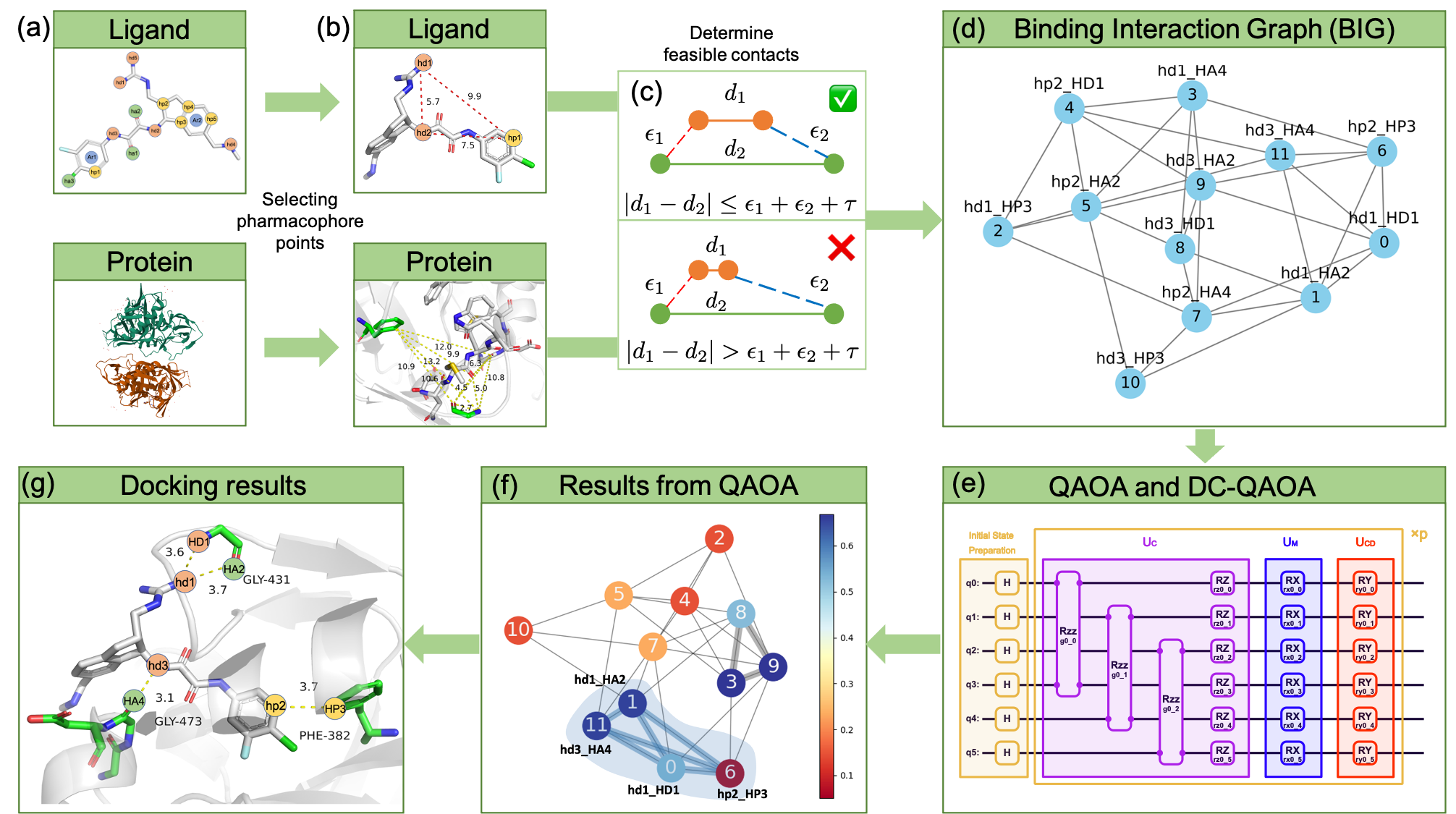} 
    \caption{{Illustration of the protein-ligand docking processes with employing the QAOA and DC-QAOA, exemplified by the complex formation involving HIV-1 gp120 and JP-III-048 (PDB ID: 5F4L). Panel (a) displays the structures of the protein and ligand. For analytical purposes, three pharmacophores on the ligand and four on the protein were identified, leading to the construction of LDGs for each, as depicted in Panel (b). Subsequently, the BIG is defined, its vertices quantitatively correlating to the product of the number of selected pharmacophores on both ligand and protein. The establishment of edges between any two vertices follows the criteria illustrated in Panel (c). Panel (d) illustrates the resultant graph. The application of both the QAOA and its derivative, DC-QAOA algorithms, the latter incorporating CD driving terms, is then implemented. The results of these computational solutions are presented in Panel (f), with the corresponding docking poses elucidated in Panel (g).}}
    %}
\label{fig:process}
\end{figure*}

\section{\label{sec:Methods} Methods}

{The exploration of drug discovery often involves the identification of small molecules, or ligands, that demonstrate high-affinity binding to specific target proteins. This binding can modulate the activity, functionality, or interactions of the protein with other biomolecules, thus eliciting therapeutic effects.} However, the conformational flexibility of proteins and ligands, coupled with the vast search space of potential binding poses, makes this a computationally challenging problem. To address this challenge, we first map the molecular docking problem onto the maximum vertex weight clique problem, as previously outlined in Ref.~\cite{kuhl1984combinatorial}. {This approach is particularly intriguing due to its potential applicability in quantum computing, notably in conjunction with Gaussian boson sampling as explored in earlier studies.}~\cite{banchi2020molecular,yu2022universal}. {However, the exploration of molecular docking problem on general-purpose quantum computers, particularly those that are gate-based or circuit-based, remains limited.} Here, we study alternative ways to solve molecular docking using digital circuit-based quantum computers. 

\subsection{Mapping molecular docking to maximum vertex weight clique}
% 
%\replaced{In the realm of protein-ligand docking, the significance of any atoms is not uniformly distributed. Our focus is primarily on pharmacophores — specific sets of atoms that play a pivotal role in dictating a molecule's pharmacological and biological interactions.}{In protein-ligand docking, not all atoms are of equal importance. Instead, we focus on pharmacophores~---~specific sets of points that significantly influence a molecule's pharmacological and biological interactions.}\replaced{To represent the geometric relationship of pharmacophores on proteins(ligands), we create a labeled distance graph(LDG).}{To articulate the distance relationship between these pharmacophore points in proteins or ligands} \replaced{In the construction of the LDG for proteins and ligands, vertices represent selected pharmacophore points, with each vertex distinctly annotated to indicate its typology. The vertices are interconnected by edges, whose weights are determined by the Euclidean distances between corresponding pharmacophore points.}{For the protein or ligand molecule under investigation, we can select pharmacophore points involved in binding interactions to form the vertices, each labeled according to its type. Edges connect these vertices, with weights assigned based on the Euclidean distance between the corresponding pharmacophore points.}

{In the realm of protein-ligand docking, the significance of any atoms is not uniformly distributed. Thus, we focus on pharmacophores--specific sets of atoms that play a pivotal role in dictating a molecule's pharmacological and biological interactions.} These pharmacophores include elements with negative or positive charges, the hydrogen-bond donor or acceptor, the hydrophobe, and the aromatic ring. {Given the constraints imposed by the current limitations of quantum devices, our study strategically narrows its focus to the most significant pharmacophores or adopts heuristic methods for selecting pharmacophore points.}

 {To represent the geometric relationship of pharmacophores on proteins/ligands, we create a labeled distance graph~(LDG). (Please refer to the Fig.~\ref{fig:process}(a-b)). In the construction of the LDG for proteins and ligands, vertices represent selected pharmacophore points, with each vertex distinctly annotated to indicate its typology. The vertices are interconnected by edges, whose weights are determined by the Euclidean distances between corresponding pharmacophore points. This graphical representation enables the elucidation of spatial relationships amongst pharmacophores, particularly focusing on the docking points of interest on the protein or ligand under study.}

After the selection and construction of the LDGs for the protein or ligand, the subsequent imperative involves the creation of a Binding Interaction Graph (BIG). The primary objective of this BIG is to serve as a visual representation of the docking relationship, thereby elucidating potential interactions among pharmacophore points.

In the BIG, the vertex, denoted as $(v_{l}, v_{p})$, encapsulates a pair of points, where $v_{l}$ denotes a vertex originating from the ligand, and $v_{p}$ signifies a vertex from the protein. Consequently, the total vertex count in the BIG is symbolized as $N = nm$, with $n$ and $m$ indicating the number of vertices in the LDG of the ligand and the protein, respectively. This integrative process results in a holistic representation, encompassing all possible theoretical interaction pairings, herein termed \textit{connects}. Within this framework, each edge in the BIG, expressed as $(v_{l_{i}}, v_{p_{i}}) - (v_{l_{j}}, v_{p_{j}})$, delineates two vertices on the edge that are capable of co-existing in a potential docking posture. The presence of a clique in the BIG implies that every vertex within the clique can concurrently exist, thereby depicting a feasible docking posture. Furthermore, the resolution of the MVWCP within the BIG identifies the most probable docking posture from all potential configurations.

%This pivotal step is elucidated in detail in Fig.~\ref{fig:process}(d). The BIG is constructed through the integration of the LDG of both the ligand and the protein. 

%This notation facilitates the possibility of selecting any vertex from the set of $n$ vertices on the ligand and any from the set of $m$ vertices on the protein, thereby constituting a vertex within the BIG.
%\replaced{Nevertheless, owing to the geometric constraints inherent in the ligand and its binding site, not all combinations of contact points are practically feasible.}{The binding interaction graph serves to represent potential binding poses via contact sets. However, due to the geometric constraints of the ligand and binding site, not all contact combinations are feasible.} Contact compatibility is determined by the geometric distances between contacts on the ligand and the binding site, with a flexible contact pair forming when the distances between the corresponding pharmacophore points do not exceed $\tau + 2 \epsilon$, where $\tau$ and $\epsilon$ denote the flexibility constant and interaction distance, respectively [Please see FIG.~\ref{fig:process}(c)]. (refer to FIG.~\ref{fig:process}(c) for further elucidation).}  
%Subsequently, we exclusively establish edges between these compatible contacts, thereby constructing a complete subgraph that represents a feasible binding pose. 
%: one located on the ligand and the other on the protein, 

The inherent geometrical constraints of both the ligand and its binding site fundamentally limit the feasibility of certain combinations of contact points. This necessitates an initial assessment aimed at identifying contact points that are mutually compatible. The primary criterion for determining the compatibility of these contact points hinges on the geometric proximities between the ligand's and the binding site's respective contact points.

In this analysis, compatibility of a pair of contact points is determined based on the absolute difference in pharmacophore point distances, \( |d_1 - d_2| \). This difference must meet a specific threshold defined as the sum of the flexibility constant \( \tau \) and interaction distances \( \epsilon_1 \) and \( \epsilon_2 \). As depicted in FIG.~\ref{fig:process}(c), a pair is deemed compatible if \( |d_1 - d_2|  \geq \tau + \epsilon_1 + \epsilon_2\); otherwise, it is incompatible. Here, \( d_1 \) is the Euclidean distance between ligands at vertices \( v_{l_{1}} \) and \( v_{l_{2}} \), and \( d_2 \) is the distance between proteins at vertices \( v_{p_{1}} \) and \( v_{p_{2}} \). The parameters \( \tau \), \( \epsilon_1 \), and \( \epsilon_2 \) represent the flexibility allowance and allowed interaction distances.

Choosing a specific $\epsilon$ value involves combining empirical observations, experimental data, and theory. This choice depends on the pharmacophore's nature at both interaction ends, as different pharmacophores create unique interactions with their own typical distances. For example, hydrogen bonding occurs when a hydrogen atom in the ligand, attached to an electronegative atom like oxygen or nitrogen, forms a favorable interaction with a hydrogen bond donor or acceptor in the protein. These bond lengths are usually between 2.5 to 3.3 \AA. Hydrophobic interactions happen between non-polar groups in the ligand and hydrophobic protein residues, caused by water molecules being excluded from the hydrophobic area. The ideal distances for these interactions are generally between 3.3 to 4 \AA. $\pi-\pi$ stacking happens when aromatic rings in the ligand and protein align, forming interactions through $\pi$-electron clouds, with typical distances around 3.5 to 4.5 \AA.

Understanding these interactions leads to a flexible approach in selecting $\epsilon$. If all pharmacophore points are hydrogen bond receptors and donors, choosing a shorter distance can improve docking prediction accuracy. However, if the pharmacophore points involve different interaction types, selecting a fixed interaction distance can simplify the process. This adaptable method enhances docking accuracy by adjusting to various pharmacophore point interactions. The effectiveness of this approach will be shown in the upcoming examples in Sec.~\ref{sec:Examples}.

% {Generally, as illustrated in FIG.~\ref{fig:process}(c-d), a systematic approach is employed to ascertain the existence of an edge within the BIG. Each vertex in the BIG encapsulates a pair of points: one located on the ligand and the other on the protein. The parameter $d_1$ denotes the Euclidean distance between the ligands at the two vertices connected by an edge. Similarly, $d_2$ signifies the Euclidean distance between the proteins at these vertices. An edge is deemed compatible if the condition $d_{2} \geq + \epsilon_{1} +\epsilon_{1} +\tau$ is met. This criterion ensures a precise and reliable determination of compatibility between vertices. In instances where this condition is not satisfied, the vertices are classified as incompatible, indicating a lack of feasible binding interaction.}

\subsection{Mathematical optimization formulations of maximum vertex weight clique }
%According to Eq.\ref{eq:mvwc2}, any violated constraint, i.e., for each $\{v_i, v_j\} \in \bar{E}$, adds a penalty value of $2P$ to the objective value in a solution.

%\replaced{Considering a specific molecular docking problem, the BIG is constructed, denoted as $\text{BIG} =(V, E)$, where $V$ represents the set of vertices and $E$ signifies the feasible set of edges. In this graph, each vertex is associated with a positive weight $w_i$, indicative of its pharmacophore potential within the context of molecular docking. The MVWCP in this context is formulated as a Binary Quadratic Programming (BQP) problem, as delineated in \cite{wang2016solving}:}{Suppose we have obtained the BIG for a specific molecular docking problem. Consider the undirected graph $\text{BIG} =(V, E)$, where $V$ and $E$ represent the vertex and edge sets, respectively. Each vertex is assigned a positive weight $w_i$, which denotes the pharmacophore potential in molecular docking problem. The MVWCP can be modeled as a binary quadratic programming (BQP) problem as follows~\cite{wang2016solving}:}

Considering a specific molecular docking problem, the BIG is constructed, denoted as $\text{BIG} =(V, E)$, where $V$ represents the set of vertices and $E$ signifies the feasible set of edges. In this graph, each vertex is associated with a positive weight $w_i$, indicative of its pharmacophore potential within the context of molecular docking. The MVWCP in this context is formulated as a Binary Quadratic Programming (BQP) problem, as delineated in \cite{wang2016solving}:
\begin{equation}
\begin{aligned}
\operatorname{maximize} \quad & \sum_{i=1}^N w_i x_i+\sum_{i=1}^N \sum_{j=1, j \neq i}^N w_{i j} x_i x_j, \\
\text{subject to:} \quad & x_i \in\{0,1\}, \quad \forall i \in\{1, \ldots, N\}.
\end{aligned}
\label{eq:mvwc2}
\end{equation}

Here, $N=|V|$ denotes the scale of the problem, equating to the number of qubits utilized for its resolution. The binary variable $x_i$ corresponds to the vertex $v_i$. The weight $w_{i j}=P$ is assigned if $\left\{v_i, v_j\right\} \in \bar{E}$, and zero otherwise, where $\bar{E}$ represents the edge set of the complementary graph $\bar{G}$, and $P$ is a pre-determined negative penalty scalar. This BQP formulation, a nonlinear interpretation of MVWCP, parallels prior investigations, such as those outlined in \cite{horst2000introduction}. The quadratic function yields an equivalent objective value to the linear form when all penalty terms are set to 0, signifying the fulfillment of all constraints. Theoretically, a higher absolute magnitude of $|P|$ simplifies attaining accurate results, as it deters the formation of non-permissible edges. However, empirical observations from quantum computational simulations suggest a non-linear relationship with $P$, indicating that excessive values can lead to sub-optimal results. This can be attributed to the disproportionate reduction of $w_i$'s coefficient after normalization, which impedes the efficacy of QAOA in finding the optimal solution. These phenomena are further elucidated through subsequent examples in Sec.~\ref{sec:Examples}.

%$\bar{E}$ denotes the edge set of the complementary graph $\bar{G}$, $w_{i j}=P$ if $\left\{v_i, v_j\right\} \in \bar{E}$ and 0 otherwise, and $P$ is a negative penalty scalar. In scenarios involving classical computation, it is common to set $P \geq -1000 \sum_{i=1}^{N} w_i$. According to Eq.~(\ref{eq:mvwc2}), violation of any constraint, particularly $\left\{v_i, v_j\right\} \in \bar{E}$, incurs an additional penalty of $2P$ to the objective. 

Notably, if $w_i = 1$ for all $i \in\{1, \ldots, N\}$, Eq.~\eqref{eq:mvwc2} transforms into the linear model of the classical maximum clique problem (MCP). The MCP, as a decision problem, is established in the literature as NP-complete, presenting significant computational challenges~\cite{garey1979computers}. The MVWCP, as a more complex variant of the MCP, inherently poses computational obstacles that are at least as formidable. Consequently, efficiently and precisely solving an arbitrary MCP is improbable. Nonetheless, in the context of approximate solutions, quantum computing holds the potential for enhanced efficiency, as evidenced by prior research in graph theory concerning the maximum cut (MaxCut) and maximum independent set (MIS) problems~\cite{farhi2014quantum,basso2021quantum,wang2018quantum,choi2019tutorial,zhou2020quantum,wurtz2021maxcut,marwaha2021local,barak2021classical,farhi2020quantum,farhi2020quantum}, in the realms of the Sherrington-Kirkpatrick (SK) model and the $k$-spin model within the theory of spin glasses and disordered systems~\cite{farhi2022quantum,bandeira2019computational,montanari2021optimization,el2021optimization,basso2022performance}, the problem of Low Autocorrelation Binary Sequences (LABS)~\cite{ shaydulin_evidence_2023}, and, more broadly, in quadratic unconstrained binary optimization (QUBO) problems~\cite{lucas2014ising} (see also the review~\cite{blekos2023review}). Gratifyingly, the optimization problem represented by Eq.~(\ref{eq:mvwc2}) can also be formulated as a QUBO problem, rendering it especially amenable to quantum computing approaches.

To construct the corresponding Hamiltonian, we map the binary variables $x_i$ to $\sigma_{i}^{z}$ matrices acting on individual qubits. Given the relationship as $x_i = (\sigma_{i}^{z} -1 ) / 2$, the objective function can be rewritten in terms of $\sigma_{i}^{z}$:
\begin{equation}\label{eq:loss}
    H= \frac{1}{2} \sum_{i \in V} w_i\left(\sigma_{i}^{z} - 1\right)+\frac{P}{4} \sum_{(i, j) \notin E, i \neq j}\left(\sigma_{i}^{z} -1 \right)(\sigma_{j}^{z}-1).
\end{equation}
Simplifying the above Hamiltonian yields:
\begin{equation}
    H_C = \sum_{i \in V} w_i \sigma_{i}^{z} +\frac{P}{2} \sum_{(i, j) \notin E, i \neq j}\left(\sigma_{i}^{z} \sigma_{j}^{z}-\sigma_{i}^{z}-\sigma_{j}^{z}\right) + const.
\label{eq:Hamiltonian}
\end{equation}
Here, the constant term does not influence the optimization process and can be omitted during the implementation of the quantum algorithm. By minimizing the expectation value of the cost Hamiltonian $H_C$ with respect to a quantum state, we can find the ground state, which corresponds to the largest weighted clique in the graph. 

%Compared to general VQAs~\cite{cerezo2021variational}, QAOA offers certain advantages and characteristics that make it particularly well-suited for tackling combinatorial optimization problems~\cite{farhi2014quantum}. QAOA has a simple and interpretable structure, making it easier to scale up and analyze. It consists of repeated applications of two types of quantum gates: the problem-dependent mixing operator and the problem-independent cost operator. This simplicity enables straightforward parameter tuning and facilitates the identification of optimal solutions~\cite{farhi2014quantum,zhou2020quantum}. It is also designed to naturally capture the structure of combinatorial optimization problems. By utilizing a cost operator that encodes the problem's objective function and a mixing operator that explores the solution space, QAOA can exploit the problem's underlying structure efficiently. Thus, QAOA provides a powerful tool to tackle the above combinatorial problem in molecular docking.

\subsection{\label{sec:QAOA} QAOA and DC-QAOA}

The QAOA is a quantum algorithm tailored for solving combinatorial optimization problems. This method exploits the adiabatic principle, facilitated by a time-dependent Hamiltonian $H_{\text{adi}}(t)$, defined as:
\begin{equation}
H_{\text{adi}}(t) = [1 - \lambda(t)] H_{\text{M}} + \lambda(t) H_{\text{C}},
\end{equation}
where $\lambda(t)$, a monotonic function within the range [0,1], orchestrates the annealing protocol over the interval $t \in [0, T]$. The algorithm initializes the system in the ground state of a simpler Hamiltonian $H_{\text{M}}$ and then adiabatically evolves towards the problem Hamiltonian $H_{\text{C}}$, whose ground state encodes the desired solution. According to the adiabatic theorem, if the evolution is sufficiently slow, particularly ensuring a non-vanishing energy gap between the ground state and the first excited state, the system predominantly remains in the ground state. The solution is subsequently extrapolated from the ground state of $H_{\text{C}}$. 

The QAOA approximates this evolution by alternating unitary operations $U_M(\beta_j) = \exp(-i \beta_j H_{\text{M}})$ and $U_C(\gamma_j) = \exp(-i \gamma_j H_{\text{C}})$, with parameters $\beta_j$ and $\gamma_j$. The process, limited to $p$ iterations, is described as:
\begin{equation}
   \begin{aligned}
U(\boldsymbol{\gamma}, \boldsymbol{\beta}) = & U_M\left(\beta_p\right) U_C\left(\gamma_p\right) U_M\left(\beta_{p-1}\right) U_C\left(\gamma_{p-1}\right) \ldots \\
& \ldots U_M\left(\beta_1\right) U_C\left(\gamma_1\right),
\end{aligned}
\end{equation}
Here, $p$ denotes the number of alternating applications of $U_C$ and $U_M$, we also called the layers of QAOA. It is evident that the search space expanded by $p$ layers of unitary operators $U_C$ and $U_M$ is more extensive than that of $p - 1$ layers.
% , with the QAOA ansatz depth characterized by the $2p$ parameter space $(\boldsymbol{\gamma}, \boldsymbol{\beta})$.

The expected value of Hamiltonian $H_C$ in the quantum state is given by $E_p(\boldsymbol{\gamma}, \boldsymbol{\beta})$:
\begin{equation}
E_p(\boldsymbol{\gamma}, \boldsymbol{\beta}) = \langle \boldsymbol{\gamma}, \boldsymbol{\beta} | H_C | \boldsymbol{\gamma}, \boldsymbol{\beta} \rangle.
\end{equation}

The conventional QAOA commonly uses a single global parameter for $U_M(\beta)$ and $U_C(\gamma)$. To achieve better results, a significant number of layers $p$ are often necessary. However, a recent enhancement known as the Multi-angle QAOA (ma-QAOA), or Full-Parameter QAOA (FP-QAOA), has been proposed~\cite{herrman2022multi,shi2022multiangle}, which can effectively reduce the number of layers at the cost of increasing the parameters optimized for each layer. Unless otherwise specified, all numerical simulations presented in this paper are conducted using this more intricate and physically representative full-parameter model.

Moreover, another variant, DC-QAOA, incorporates counterdiabatic (CD) terms~(see Fig.~\ref{fig:process}(e))~\cite{chandarana2022digitized}. Studies show that this addition significantly lowers the needed value of $p$ for a near-optimal trial state~\cite{chandarana2022digitized}. Combining these benefits of two algorithms, we introduce the Multi-angle Digitized-Counterdiabatic QAOA (ma-DC-QAOA). For simplicity, we refer to it as DC-QAOA throughout our discussion.

% This approach modifies $U(\boldsymbol{\gamma}, \boldsymbol{\beta})$ to $U(\boldsymbol{\gamma}, \boldsymbol{\beta}, \boldsymbol{\alpha})$, and similarly, $E(\boldsymbol{\gamma}, \boldsymbol{\beta})$ becomes $E(\boldsymbol{\gamma}, \boldsymbol{\beta}, \boldsymbol{\alpha})$ 
%By modulating the parameters $ \boldsymbol{\gamma}, \boldsymbol{\beta} $, we achieve
%\begin{equation}
%M_p = \max_{\boldsymbol{\gamma}, \boldsymbol{\beta}} E_p(\boldsymbol{\gamma}, \boldsymbol{\beta})
%\end{equation}
%as the approximate optimal solution for a given layer count $ p $. 

% In the conventional QAOA frameworks, A higher value of $p$ typically correlates with a closer approach to the optimal solution. However, this increase in $p$ also elevates the complexity associated with the implementation of the unitary operator $U(\boldsymbol{\gamma},\boldsymbol{\beta})$. Such complexity poses substantial challenges for the NISQ devices due to efficiency concerns. 

% To address this issue, the DC-QAOA introduces counterdiabatic ($\mathrm{CD}$) terms. This integration adds an additional variational parameter, transforming $U(\boldsymbol{\gamma}, \boldsymbol{\beta})$ into $U(\boldsymbol{\gamma}, \boldsymbol{\beta}, \boldsymbol{\alpha})$ and accordingly, $E(\boldsymbol{\gamma}, \boldsymbol{\beta})$ into $E(\boldsymbol{\gamma}, \boldsymbol{\beta}, \boldsymbol{\alpha})$~\cite{chandarana2022digitized}, as depicted in Fig~\ref{fig:process}(e). Empirical studies indicate that these extra parameters significantly reduce the required value of $p$ to achieve a near-optimal trial state~\cite{chandarana2022digitized}.

The integration of the CD driving term, based on shortcuts to adiabaticity principles~\cite{chen2010shortcut,chen2010fast,hegade2021shortcuts}, mitigates nonadiabatic transitions and accelerates quantum processes, as noted in studies such as~\cite{del2012assisted,demirplak2005assisted,demirplak2003adiabatic,berry2009transitionless}, with notable applications in quantum state preparation tasks~\cite{yao2021reinforcement}. 
In this paper, we follow the approach in~\cite{chandarana2022digitized} and adopt the nested commutator framework of the adiabatic gauge potential as introduced by Claeys et al.~\cite{claeys2019floquet}, to construct a pool of CD operators. 

In the present study, we adopt the nested commutator framework for the adiabatic gauge potential, as introduced by Claeys et al.~\cite{claeys2019floquet} and further explored in Chandarana et al.~\cite{chandarana2022digitized}, which involves the construction of a comprehensive pool of CD operators. Specifically, we limit our focus to the second-order expansion within the nested commutator framework, denoted as \( l=2 \). Consequently, the operator pool \( A \) is defined as \( \{\sigma^y, \sigma^z \sigma^y, \sigma^y \sigma^z, \sigma^x \sigma^y, \sigma^y \sigma^x\} \), thereby encapsulating a range of both local and two-body interactions.

%In this approach, we confine our consideration to the second order in the expansion of the nested commutator, i.e., $ l=2 $. The operator pool $ A $ is defined as $ \{\sigma^y, \sigma^z \sigma^y, \sigma^y \sigma^z, \sigma^x \sigma^y, \sigma^y \sigma^x\} $, encompassing solely local and two-body interactions. 

Choosing the right operator pool is important and depends on the Hamiltonian being studied. Different problems may require different operators. For example, a key paper~\cite{chandarana2022digitized} presents a method for selecting counterdiabatic (CD) terms for the Ising model under various conditions. This method defines the Hamiltonian for the one-dimensional Ising spin model as $H_{\text{prob}}(\sigma) = -\sum_{(i,j)} J_{ij}\sigma^z_i \sigma^z_j - \sum_{i} h_i\sigma^z_i - \sum_{i} k_i\sigma^x_i$. In this formula, $\sigma^d_i$ represents the Pauli matrices at each site, and $(i, j)$ indicates nearest-neighbor interactions with the coupling constant $J_{ij}$.

This approach results in two scenarios. The first is the Linear Ferromagnetic Ising Model (LFIM), where $k_i = 0$ and the CD term is $A_t = \sum_{i} \sigma^y_i$. The second scenario is the Transverse-Field Ising Model (TFIM), which is defined when $h_i = 0$ and the CD term is $A_t = \sum_{i} \sigma^z_i \sigma^z_{i+1}$. In our example, we consider the Hamiltonian $H_C$ for the MVWCP as shown in Eq.~(\ref{eq:Hamiltonian}). In this case, the model matches LFIM, which leads to the CD term $A_t = \sum_{i} \sigma^y_i$.

The CD component in the quantum circuit is represented by the unitary operator
\begin{equation}
U_{\text{CD}}(\alpha) = \prod_{j=1}^{L} \exp [-i\alpha A_j^q].
\end{equation}
where $ A_j^q $ is the $ q $-local CD operator from the set $ A $. 

In circuit structure terms, DC-QAOA improves on the original QAOA by adding an extra CD term. This change leads to a deeper circuit and a more complex optimization space. In a QAOA framework where each quantum gate has its own parameter, a DC-QAOA circuit with the same number of layers will have more parameterized gates. This is especially true when the CD term is complex. However, this complexity does not mean it is less efficient. DC-QAOA can achieve its goals with fewer layers, which is beneficial for reducing the number of gates. Section 3B of this study shows this advantage. 

Moreover, the application of DC-QAOA to intricate tasks such as molecular docking is further optimized by our specific CD term, denoted as $A_t = \sum_{i} \sigma^y_i$. Adding this term only inserts some single-qubit parameter gates after each QAOA layer, but can effectively reduce the circuit depth from the subsequent numerical simulations. This reduction is crucial as it avoids the need for more complex double-qubit gates, which is an important factor considering the limitations of current quantum computing hardware. Additionally, this approach also simplifies optimization problems. Collectively, these advancements underscore the considerable potential of DC-QAOA in effectively addressing large-scale molecular docking challenges, thereby expanding its applicative scope and reinforcing its utility in the field of quantum computing.

\section{\label{sec:Examples} Numerical simulation}

% Next, we consider the application of DC-QAOA in solving three practical Protein-ligands docking problems, {The general workflow of protein-ligands docking via QAOA algorithms is illustrated in FIG.~\ref{fig:process}.}
In this section, we focus on the practical implementation of the QAOA-based algorithms in tackling three protein-ligand docking problems. The comprehensive approach used for protein-ligand docking through QAOA algorithms is illustrated in Fig.~\ref{fig:process}.

First, we demonstrate the complete process of our algorithm using a small-scale example, as illustrated in Fig.~\ref{fig:process}, and highlight the effect of flexible choice of $\epsilon$. Next, we focus on a medium-scale docking problem within an 8-qubit quantum system, and conduct a thorough analysis to evaluate the effect of the different QAOA parameters for docking process. This involved examining the number of QAOA layers ($p$), a penalty term ($P$), and evaluating the impact of quantum noise on the docking process. Finally, we applied our method to a challenging 12-qubit example, characterized by almost identical weights in the largest and second-largest cliques, which illustrates the advantages of DC-QAOA in molecular docking problems.

\subsection{The impact of the flexible $\epsilon$}

We begin our numerical investigation with a 6-qubit quantum system to explain the operational process of our proposed QAOA-based molecular docking algorithm, and focus on empirically examining the influence of the $\epsilon$ parameter on docking results. The research delineates the complex interaction between the SARS-CoV-2 main protease (Mpro), a critical target in COVID-19 therapeutics, and a covalent pyrazoline-based inhibitor, PM-2-020B. This inhibitor has been identified through the Protein Data Bank (PDB) with the ID 8SKH, as reported in the seminal work~\cite{moon2022discovery}. 

The main protease (Mpro) of the coronavirus, a crucial enzyme in viral replication, presents a potential target for therapeutic intervention. Researchers have identified and optimized pyrazoline-based covalent inhibitors for SARS-CoV-2 Mpro using activity-based protein profiling. These inhibitors have shown promising efficacy against Mpro from SARS-CoV-2 and other coronaviruses, underscoring their potential as pan-coronavirus inhibitors. In this context, we delve into the docking process between the pyrazoline-based inhibitor PM-2-020B and SARS-CoV-2 Mpro, providing a detailed investigation of this interaction.

Our method begins with the extraction of all pharmacophore points on ligands and receptors using the Python package $rdkit$~\cite{greg_landrum_2024_10460537}. We classify these points into four distinct categories: hydrogen-bond donor/acceptor (HD/HA), hydrophobe (HP), and aromatic ring (AR). Vertices on the ligand and receptor are denoted with lowercase and uppercase letters, respectively. This process yields 7 pharmacophore points on the PM-2-020B ligand and 1422 points on the SARS-CoV-2 Mpro receptor.

The QAOA-based molecular docking algorithm requires $N =m \times n$ qubits to find the maximum weighted clique, where $n$ and $m$ represent the vertex counts in the ligand and protein binding site graphs, respectively. Given the constraints of current quantum devices, it is crucial to minimize the number of molecules to be paired. To this end, we select pharmacophore points on the protein within a 5-angstrom radius of the ligand, based on the actual binding pose. While the actual binding pose remains unknown in real scenarios, a similar set of points can be procured using drug discovery knowledge. This includes heuristics selection of ligand pharmacophore points and the use of prior knowledge of binding site locations to reduce receptor pharmacophore points. Sliding windows can further minimize receptor pharmacophore points by segregating different binding site regions. As quantum devices evolve, such simplifications may become redundant.
%~\cite{madsen2022quantum}

In this example, we identified two pharmacophore points on PM-2-020B, specifically O2378 (ha1) and C2373 (hp2), as well as three pharmacophore points on SARS-CoV-2 Mpro, namely N1104 (HD1), N1108 (HD2), and N1114 (HD3). Here, the capital letter preceding the number indicates the atomic type of the Pharmacophore, while the number corresponds to the atomic serial number in the PDB file. Once we obtained the two annotated distance graphs (one for the PM-2-020B ligand and another for the SARS-CoV-2 Mpro receptor), we proceeded to construct the binding interaction graph. As previously mentioned, the vertices in this graph denote potential connectivity modes. Given our selected pharmacophores, six possible vertices emerge in the binding interaction graph: ha1-HD1, ha1-HD2, ha1-HD3, hp2-HD1, hp2-HD2, and hp2-HD3. 

\begin{figure}[htbp]
    \centering
    \includegraphics[width=0.5\textwidth]{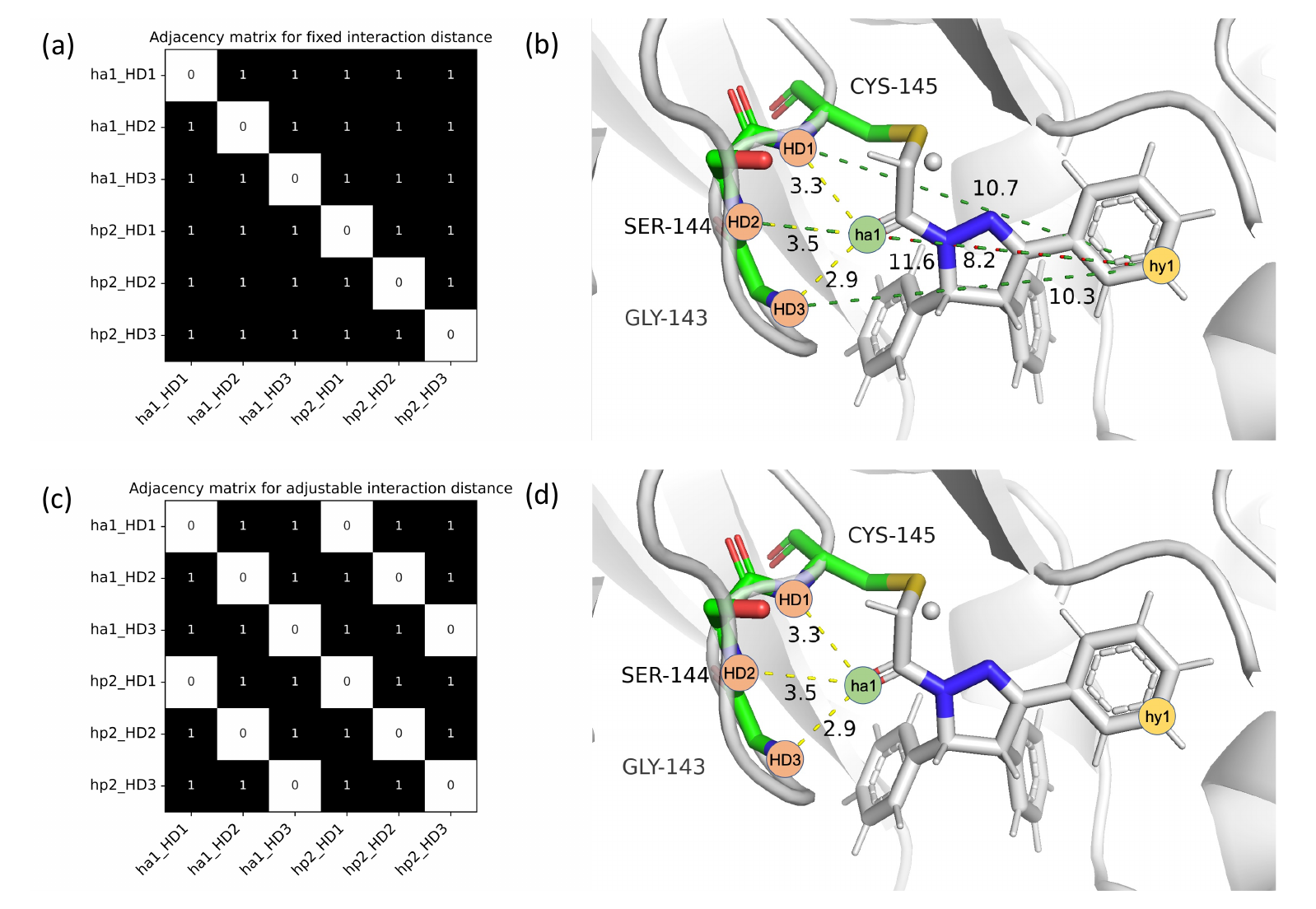} 
    \caption{The adjacency matrix corresponding to (a) a fixed distance ($\epsilon = 4$~\AA)  between all pharmacophores in the docking process. (b) illustrates the co-structure of SARS-CoV-2 Mpro (COVID-19) with covalent pyrazoline-based inhibitors PM-2-020B (PDB ID: 8SKH), assuming a fixed distance ($\epsilon$) of 4 \AA~between all pharmacophores in the docking process. (c) a flexibly setting the distance parameters for different pharmacophore types (hydrogen bond interactions are set at 3 \AA, while other interactions are set at 4 \AA). (d) demonstrates the docking results obtained by flexibly setting the distance parameters for different pharmacophore types. Figures (b) and (d) show the schematic representation of the molecular docking results, generated using the open-source software PyMOL~\cite{pymol}.}
\label{fig:8skh_interaction}
\end{figure}

The subsequent step entails assessing the coexistence of any two vertices. For this purpose, we propose two unique compatibility criteria. The first, a $\tau$ flexible approach drawn from existing literature, assumes a uniform distance $\epsilon = 4$~\AA~across all pharmacophores \cite{banchi2020molecular}. The second approach adjusts $\tau +2 \epsilon$ in the $\tau$ flexible approach to $\tau + \epsilon_1 + \epsilon_2$, assigning specific distances according to different pharmacophore types. For vertices composed exclusively of hydrogen bond acceptors and donors, we adopt a shorter interaction distance, $\epsilon_{s} $~\AA. In all other cases, the interaction distance remains at $\epsilon $~\AA. The adjacency matrix corresponding to these two scenarios is illustrated in Fig.~\ref{fig:8skh_interaction}(a) and (c), respectively.

Before solving the maximum weighted clique problem, we need to add weights to the vertices in the binding interaction graph based on the pharmacophore potential. Table.~\ref{tab:Pharmacophorepotential} shows the pharmacophore potential we applied in the numerical experiments. We use the same data as in Ref.~\cite{banchi2020molecular}, and these experience parameters are derived from the PDBbind dataset~\cite{wang2004pdbbind,wang2005pdbbind,liu2015pdb}. 

In a scenario where the interaction distance between pharmacophore points remains invariant, the resulting pharmacophore model is represented by a fully connected adjacency matrix. This is reflected in the maximum vertex weight clique, which encompasses all vertices in such a matrix. The corresponding docking results, under these conditions, are illustrated in Fig.~\ref{fig:8skh_interaction}(b). However, the situation becomes more complex when the interaction distances vary, as the adjacency matrix alone no longer provides immediate and clear insights. 

To address this issue, we construct the Hamiltonian for the modified adjacency matrix, as outlined in Eq.~(\ref{eq:Hamiltonian}), and apply both conventional QAOA and its enhanced variant, DC-QAOA, to find solutions. For detailed data on the construction of LDGs and BIG, please refer to Appendix B. 

For this study, we adopt a 3-layer QAOA circuit and configured with a penalty parameter set to \( P = 6 \), which has been empirically determined to yield optimal results. A detailed illustration of the quantum circuit is presented in Fig.~\ref{fig:8SKH_circuit}. Furthermore, the optimization process has been significantly enhanced through the integration of a Quantum natural gradient descent algorithm. The intricate details of the algorithm, including its specific parameter configurations, are comprehensively delineated in Appendix A.

The experimental methodology employed in this study involved a comprehensive evaluation of 500 distinct initial parameters. As depicted in Fig.~\ref{fig:QAOA_result}(d), a meticulous analysis of the loss curve revealed that these randomly selected parameters exerted minimal impact on the experimental outcomes. Moreover, the study delved into examining the effects of varying parameters on the probability of achieving a successful outcome. To facilitate this analysis, for each initial parameters, the quantum circuit was reconfigured using a set of optimal parameters identified through optimization. Subsequently, for the final state of each quantum circuit, we sampled 5000 times under the $Z^{\otimes N}$ basis. The success probability for the state that corresponds to the maximum weighted clique is systematically illustrated in Fig.~\ref{fig:SP}(a).

The docking results, both algorithms exhibit a remarkable consistency, particularly those involving the interactions ha1-HD1, ha1-HD2, and ha1-HD3, are depicted in Fig.~\ref{fig:8skh_interaction}(d). It is worth noting that we find that DC-QAOA only needs 80 epochs to achieve a good loss, while the QAOA algorithm needs no less than 200 iterations under the same conditions, which proves that DC-QAOA can effectively reduce the classical/quantum resources required for iterations.

Notably, when these results are juxtaposed against those obtained from the Protein-Ligand Interaction Profiler (PLIP) as referenced in~\cite{adasme2021plip}, it is evident that our calculated docking results are more congruent with established biological principles. This contrast is especially marked when compared to the results derived under the condition of fixed interaction distances between pharmacophore points. This observation underscores the significance of appropriately modulating the interaction distances between pharmacophore points in order to obtain docking results that are more in harmony with biological laws.

\begin{table}
    \begin{ruledtabular}
        \begin{tabular}{ccccc}
        \textrm{Ligand/Protein}&
        \textrm{HD}&\textrm{HA}&\textrm{HP}&\textrm{AR}\\
        \colrule
        \textrm{hd}	&0.5244	&0.6686	&0.1453	&0.1091\\
        \textrm{ha}	&0.6686	&0.5478	&0.2317	&0.0770\\
        \textrm{hp}	&0.1453	&0.2317	&0.0504	&0.0795\\
        \textrm{ar}	&0.1091	&0.0770	&0.0795	&0.1943\\
        \end{tabular}
    \end{ruledtabular}
    \caption{Knowledge-based pharmacophore potential. Data is derived from the PDBbind dataset from Ref.~\cite{wang2004pdbbind,wang2005pdbbind,liu2015pdb}. HD: hydrogen-bond donor; HA: hydrogen-bond acceptor; Hp: hydrophobe; AR: aromatic.}
    \label{tab:Pharmacophorepotential}
\end{table}

% \subsection{8 qubits--PDB ID: 3HAC}

\subsection{Exploring the optimal parameter settings for QAOA-based Algorithm}

Next, we begin a detailed analysis assessing the impact of various QAOA parameters on the docking process. We chose an 8-qubit system for its more complex solution space compared to a 6-qubit system, while maintaining effective simulation across multiple trials.
% we examine the application of the QAOA-based molecular docking algorithm in an 8-qubit quantum system, with a focus on the structure of Dipeptidyl Peptidase-4 (DPP-4) complexed with piperidine-fused imidazopyridine 34 (PDB ID: 3HAC). Serving as a medium-scale example, our study provides an in-depth analysis of how various QAOA parameters and noise models affect the results.}
%We examine the application of the QAOA-based molecular docking algorithm in an 8-qubit quantum system, with a focus on the structure of Dipeptidyl Peptidase-4 (DPP-4) complexed with piperidine-fused imidazopyridine 34 (PDB ID: 3HAC). 

We focus on the structure of Dipeptidyl Peptidase-4 (DPP-4) complexed with piperidine-fused imidazopyridine 34 (PDB ID: 3HAC). Piperidine-fused imidazopyridine 34 has been identified as a potent inhibitor of DPP-4 and shows promise in enhancing glycemic control \cite{edmondson2009aminopiperidine}. Understanding its interactions with DPP-4 is vital for elucidating the drug's mechanism of action, optimizing its structure for increased efficacy, and potentially revealing new treatments for diabetes. This research is pivotal in advancing more effective therapies for diabetes management.

%\replaced{Unlike the previous example with six qubits, this example concentrates on a docking problem involving an aromatic ring.}{This example focuses on the docking problem that involves an aromatic ring.} 
Unlike the previous example with six qubits, this example concentrates on a docking problem involving an aromatic ring. We selected two pharmacophore points on the piperidine fused imidazopyridine 34 ligand: N12298 (hd1), and the pseudo-atoms at the center of the aromatic ring (ar2, the central coordinates of six atoms: C12301, C12302, C12306, N12307, C12308, and C12309). We also selected four pharmacophore points in DPP-4: 01453 (HA1), O1462 (HA2), C5171 (HA3), and the pseudo-atoms in the center of the aromatic ring (AR4, the central coordinates of six atoms: C2683, C2684, C2685, C2686, C2687, and C2688). We then constructed the binding interaction graph, the adjacency matrix of which is depicted in Fig.~\ref{fig:3hac_interaction}(a) (Detailed calculations on LDGs and BIG can be found in Appendix C).

\begin{figure*}[htbp]
    \centering
    \subfigure[]{\includegraphics[width=0.48\textwidth]{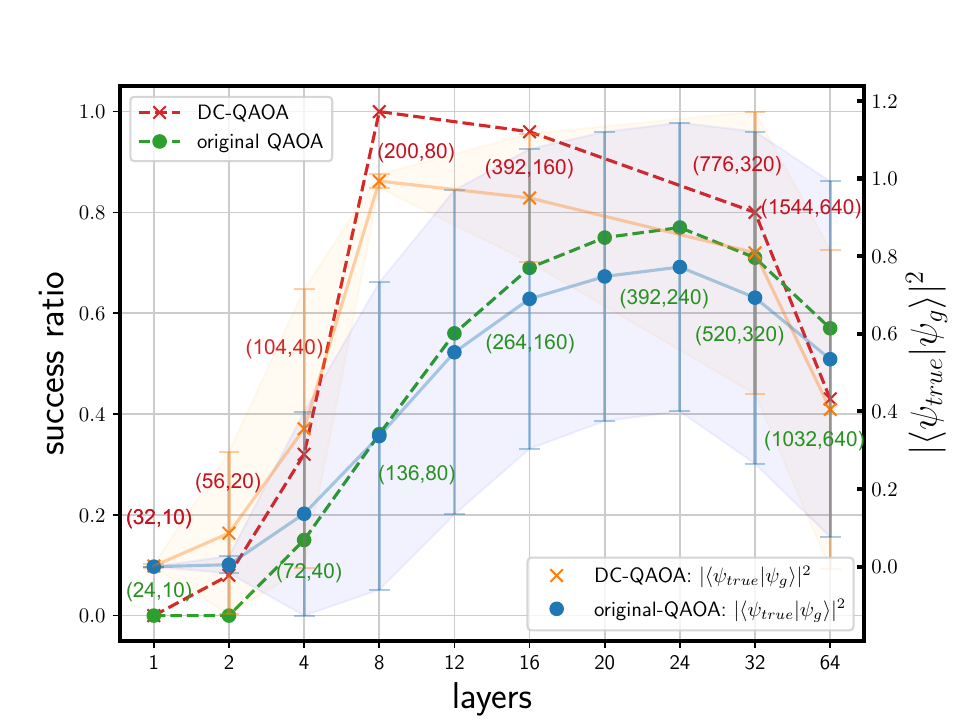}} 
    \subfigure[]{\includegraphics[width=0.48\textwidth]{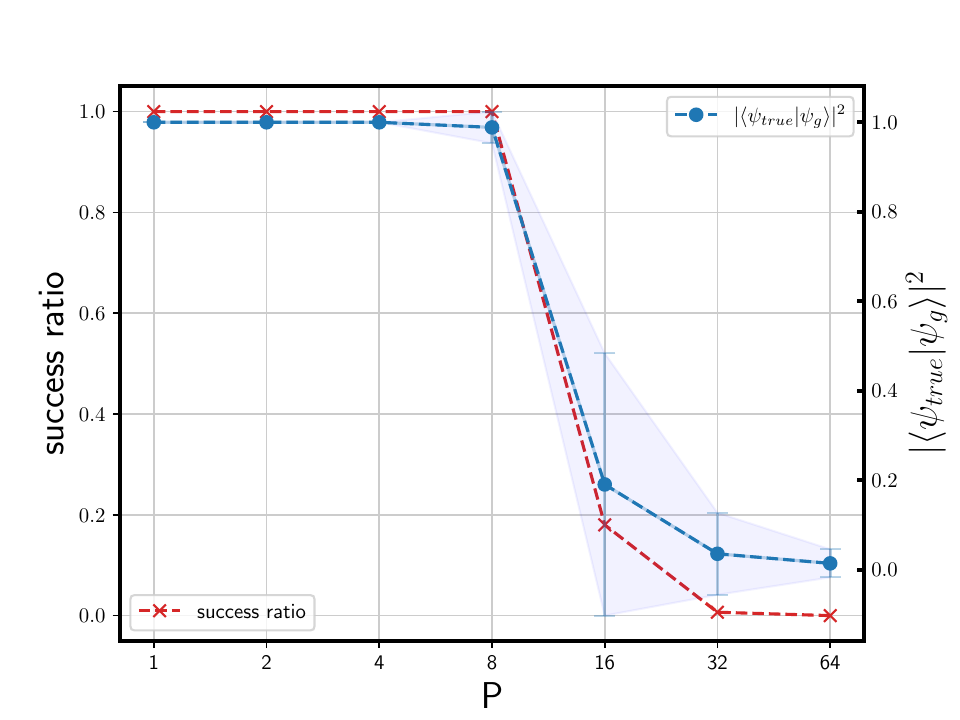}} 
    \caption{Comparative analysis of the original QAOA and DC-QAOA performance with varying numbers of layers and penalty term.  The left panel delineates the success ratio $P_{\text{true}}$ and the expected probability amplitude $|\braket{\psi_{\text{true}}|\psi_{g}}|^2$ as a function of the quantum circuit layers, depicted on the x-axis. The success ratio $P_{\text{true}}$ for the original QAOA is denoted by a green dotted line with circular nodes, whereas the DC-QAOA's is represented by a red dotted line with 'x' nodes. The expected probability amplitude $|\braket{\psi_{\text{true}}|\psi_{g}}|^2$ for the original QAOA is denoted by a solid yellow line, whereas the DC-QAOA's is represented by a solid blue line, and the shaded regions reflect the statistical error margins computed from a set of 100 samples. The annotated numerical pairs (red for DC-QAOA and green for QAOA) enumerate the quantum gate counts at varied layer depths, with the initial numeral indicating single-qubit gate counts and the trailing numeral specifying two-qubit gate counts. The right graph offers a focused view on the diminishing trend of success ratios $P_{\text{true}}$ and probability amplitudes $|\braket{\psi_{\text{true}}|\psi_{g}}|^2$ for DC-QAOA as the penalty term \( P \) increases.}
    \label{fig:diff_p_layer}
\end{figure*}

\begin{figure}[htbp]
    \centering
    \includegraphics[width=\linewidth]{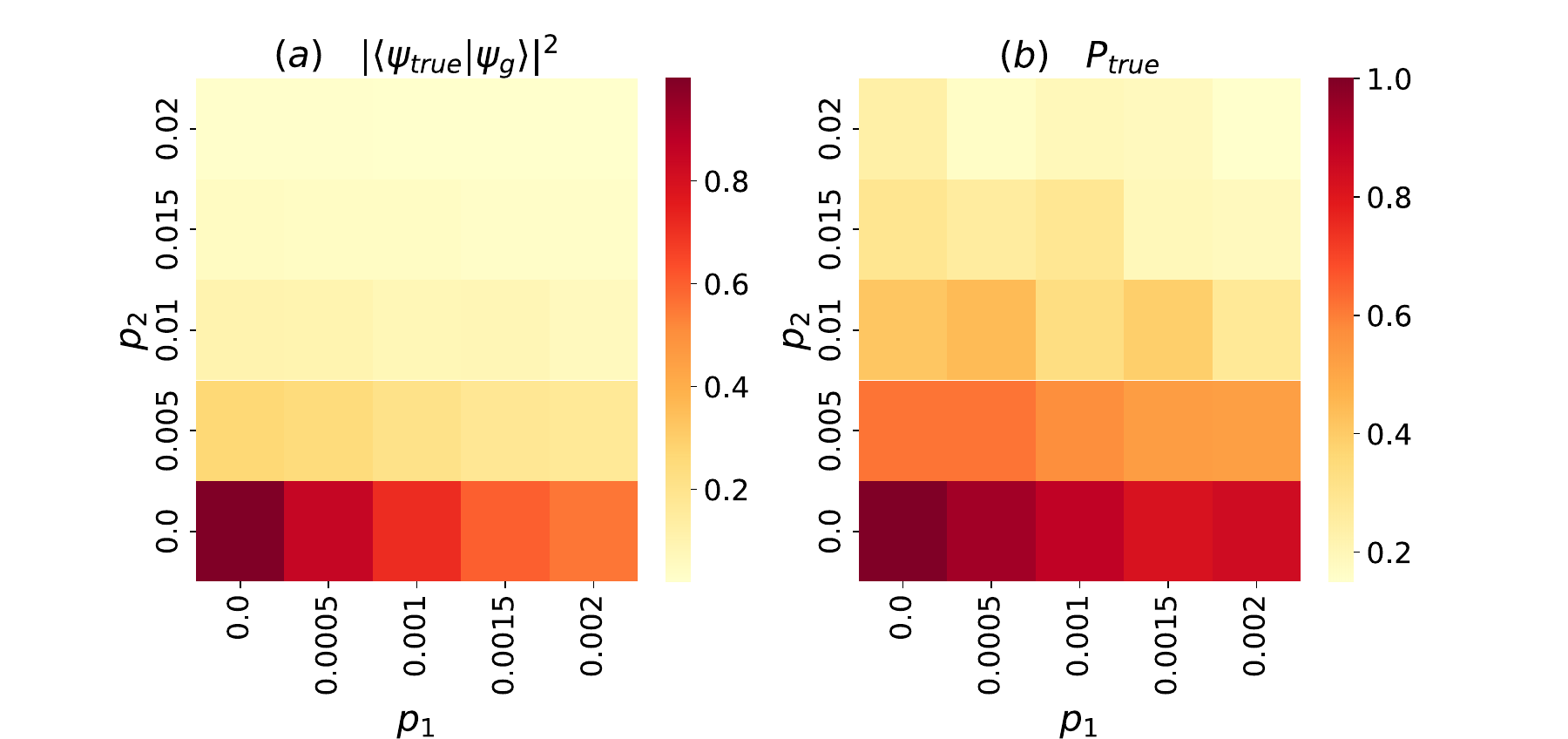}
    \caption{Panel~(a) and (b) represents the performance of DC-QAOA across different noise level, where \( p_1 \) and \( p_2 \) refer to the depolarizing noise level of single and two qubit gates respectively. Panel~(a) shows the average overlap between the optimal states and DC-QAOA output states, i.e. $|\braket{\psi_{true}|{\psi_{g}}}|^2$, in each block. Panel~(b) the average success ratio of finding optimal solution over 500 runs.}
    \label{fig:noise}
\end{figure}

%and the other is the overlapping of the target state and the Hamiltonian ground state a significant enhancement in the success probability, denoted as $P_{\text{true}}$, which represents the probability of the post-sampling string aligning with the target state $\ket{\psi_{\text{true}}} = 10000111$. Here, the green and red dotted lines represent the QAOA and DC-QAOA results, respectively. Further, our study delves into the expected probability amplitude $|\braket{\psi_{\text{true}}|\psi_{g}}|^2$ for the state $\ket{\psi_{\text{true}}} = 10000111$ Here psi_g represents the ground state of the problem Hamiltonian Hc.  indicate that the probability of success rises rapidly and then decreases as the number of layers increases In the graphical representation, the solid blue and yellow lines depict the performance metrics for QAOA and DC-QAOA, while the shaded area indicates the error margin after 100 times sampling. 
% In this study, we employed two distinct metrics to elucidate the effectiveness of our approach. The first metric, denoted as \( P_{\text{true}} \), quantifies the success probability of the algorithm. Specifically, it measures the likelihood that the post-sampling string aligns with the highest probability state, corresponding to the target state \( \ket{\psi_{\text{true}}} \). The second metric assesses the degree of overlap between the target state and the Hamiltonian ground state \( \psi_{g} \), expressed as \( |\braket{\psi_{\text{true}}|\psi_{g}}|^2 \).

In this example, we also implemented both the conventional QAOA and DC-QAOA, and the quantum circuit diagram is shown in Fig.\ref{fig:3HAC_circuit} in Appendix C. This investigation rigorously examines the disparities between these two algorithms and the impact of various parameters on the performance of algprithms. Key influencing factors considered include the layers $p$, the magnitude of the penalty term $P$, and the performance of algorithm with in noise model.

To rigorously evaluate the efficacy of our algorithm, we adopted two distinct metrics. The first metric, designated as \( P_{\text{true}} \), quantifies the success ratio of the algorithm. Specifically, it measures the likelihood that the post-sampling string corresponding to the highest probability state aligns with the target state \( \ket{\psi_{\text{true}}} \). The second metric assesses the overlap between the target state and the ground state \( \psi_{g} \) of problem Hamiltonian $H_c$ , expressed as \( |\braket{\psi_{\text{true}}|\psi_{g}}|^2 \).

%In parallel, the solid blue and yellow lines represent the \( |\braket{\psi_{\text{true}}|\psi_{g}}|^2 \) values for the two algorithms, with the shaded areas indicating the error margins after 500 iterations of initial parameter selections in the quantum circuit. This trend is depicted through the graphical representation, where the green and red dotted lines correspond to the \( P_{\text{true}} \) values for QAOA and DC-QAOA, respectively.

Our analysis initially focused on evaluating the impact of the number of layers within QAOA and DC-QAOA, on their respective performance results. We noted that both \( P_{\text{true}} \) and \( |\braket{\psi_{\text{true}}|\psi_{g}}|^2 \) initially rose and then fell as more layers were added to the quantum circuit, as shown in Fig.~\ref{fig:diff_p_layer}(a). This trend is in line with theoretical predictions, suggesting that adding layers might mimic adiabatic quantum computing, improving results. However, too many layers add complexity and increase the risk of hitting local minima in the optimization process. Our findings imply that a balanced approach—increasing layers linearly with system size, following an \( O(n) \) scale—may be most effective. This strategy balances efficiency and effectiveness, reducing the risks of too many layers. Thus, it appears as a valuable method for enhancing quantum algorithm performance in computational tasks.

This analysis categorically establishes the differences in resource allocation between DC-QAOA and conventional QAOA, focusing particularly on gate requirements. Importantly, integrating selected CD terms into DC-QAOA does not require extra complex two-qubit gates, thus bolstering its experimental applicability. At \( p=8 \) layers, DC-QAOA reaches its optimal efficacy, employing only 200 single-qubit and 80 two-qubit parametric gates. Conversely, conventional QAOA attains its peak performance at a significantly higher layer of \( p=24 \), necessitating 392 single-qubit and 240 two-qubit parametric gates. Moreover, the fidelity metric \( |\braket{\psi_{\text{true}} | \psi_{g}}|^2 \) in QAOA achieves a maximum near 0.8, reflecting its reduced effectiveness. Notably, the maximum iterations in our simulation are limited to 500, as DC-QAOA achieves satisfactory results within this range. This comparative study illustrates that conventional QAOA requires twice as many single-qubit and three times as many two-qubit parametric gates as DC-QAOA. This pronounced disparity in gate utilization highlights not only superior efficiency of DC-QAOA but also suggests a heightened susceptibility of conventional QAOA to noise disturbances in quantum systems, attributed to its greater dependence on quantum gate complexity.

In the second analysis, we investigated the impact of varying the penalty term, \( P \), on docking results, as illustrated in Fig.~\ref{fig:diff_p_layer}(b). Adjustments in  \( P \) to values 1, 2, 4, or 8 yielded no significant alteration in success probability. However, when \( P \) values exceed 8, a significant decrease is evident in both \( P_{\text{true}} \) and \( |\langle \psi_{\text{true}}|\psi_{g} \rangle|^2 \). This decline can be attributed to the increased penalty term, which diminishes the contribution of the dominant term in Eq.~(\ref{eq:Hamiltonian}). As a result, it obscures the differentiation between optimal and sub-optimal outcomes, consequently increasing the probability of errors.

Next, we investigated the impact of noise models on the effectiveness of docking. We introduced varying levels of depolarizing noise into the quantum circuit, as illustrated in FIG.~\ref{fig:noise}. Our experimental configuration utilized an eight-layer quantum circuit with an associated penalty term set to eight. The presented results are the product of an extensive series of 500 trials for each data point, with randomized parameter initialization in each trial. Our analysis reveals a clear decrease in both the success rate of docking and the probability amplitude, directly correlated with the increased intensity of noise.

% {Next, we examined the influence of noise models on the docking efficacy. Depolarizing noise of varying strengths was incorporated throughout the quantum circuit, and the results are depicted in Fig.~\ref{fig:noise}. Our experimental setup employed a quantum circuit comprising eight layers, and a penalty term was set at eight. The results presented herein are the culmination of an extensive series of 500 trials for each data point, each trial featuring randomly initialized parameters. Our analysis revealed a discernible decline in both the docking success rate and the probability amplitude, which can be directly attributed to the increment in noise intensity.}

%The results indicate that both QAOA methods produce solutions that closely align with the ground state energy. Moreover, the results derived from the two QAOA algorithms are consistent. This congruence is depicted in FIG.~\ref{fig:QAOA_result}(a), illustrating the consistency across the two methods.

Based on the analyses conducted above, we have employed a DC-QAOA circuit featuring an eight-layer architecture and the inclusion of a penalty term set at P=8. This configuration underwent a comprehensive evaluation encompassing the initialization of 500 distinct parameter sets selected randomly. The performance of each set was meticulously monitored, and the resulting loss curves are depicted in Figure~\ref{fig:QAOA_result}(e). Remarkably, these curves manifest a significant reduction in loss error when employing the DC-QAOA. Furthermore, the DC-QAOA enables us to attain the desired outcomes within a mere 200 iterations, marking a substantial departure from the requirement of 1000 iterations for the conventional QAOA. This decrease in the iteration count translates into a noteworthy reduction in classical and quantum resource consumption during the optimization process. Subsequently, we undertake a comprehensive analysis to assess the success probability, as illustrated in Fig.\ref{fig:SP}(b). In contrast to the approach adopted in Fig.\ref{fig:diff_p_layer}, we established distinct iteration limits for DC-QAOA and QAOA: the QAOA was granted a maximum of 1500 iterations, whereas the DC-QAOA was constrained to a ceiling of 500 iterations. The DC-QAOA algorithm exhibited an exceptionally high accuracy rate, approaching 100\%, while the QAOA algorithm, although slightly less effective, still demonstrated considerable robustness in its results.

The docking outcomes, namely hd1-HA1, hd1-HA2, hd1-HA3, and ar2-AR4, have been illustrated in Figure \ref{fig:3hac_interaction}(b). It is worth highlighting that a comparative analysis performed in conjunction with data obtained from PLIP~\cite{adasme2021plip} demonstrates a notable congruence between our docking results and well-established biological principles. This concurrence not only fortifies the soundness of our methodology but also accentuates its potential to yield valuable insights within the domain of computational biology. All computational simulations were executed using the Mindquantum framework~\cite{mq_2021}.

\subsection{More qubits}

\begin{figure*}
  \centering
  \includegraphics[width=1\textwidth]{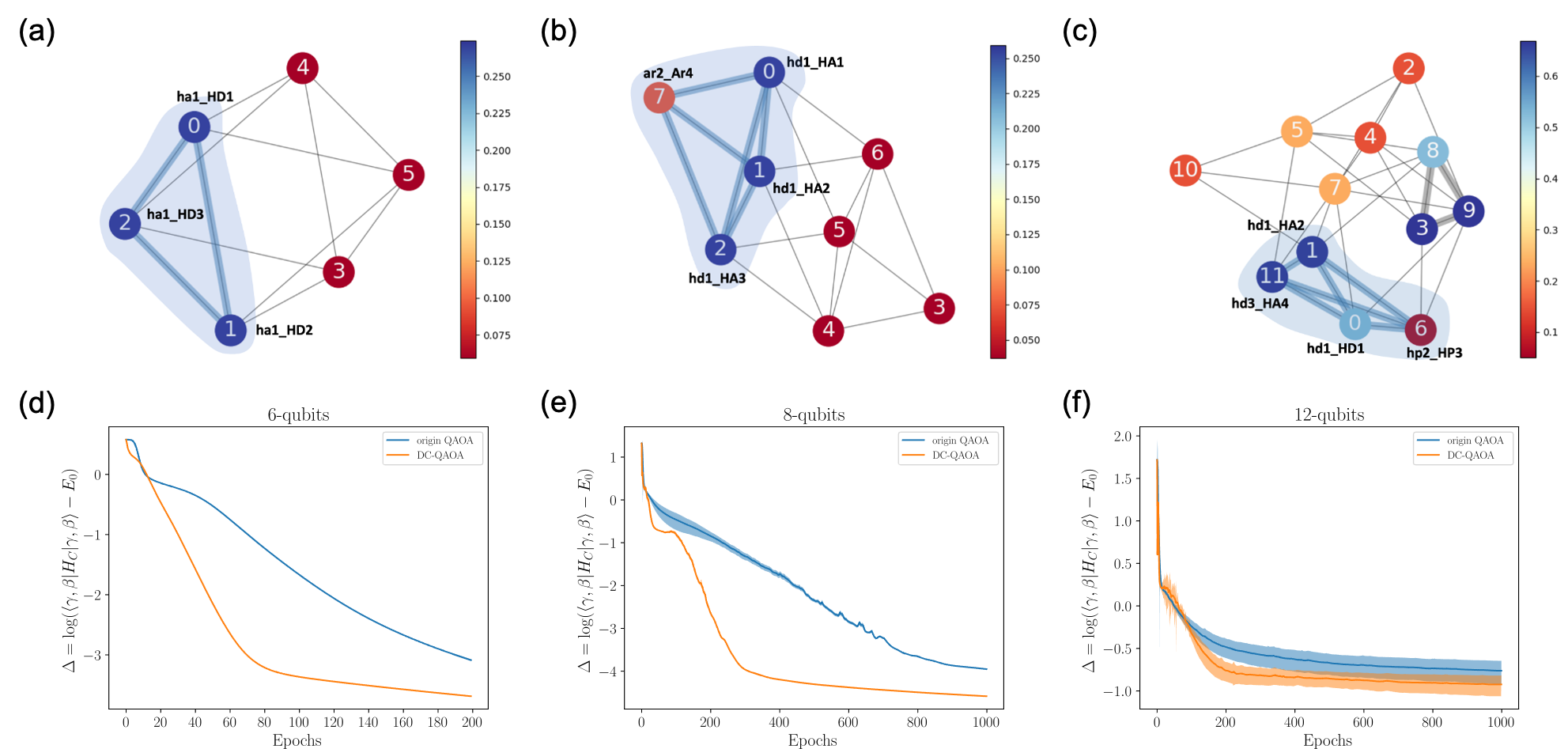}
  \caption{In Panels (a), (b), and (c), we present the results for three protein-ligand docking example form QAOA-based algorithms, specifically for PDB IDs: 8SKH (6 qubits), 3HAC (8 qubits), and 5F4L (12 qubits). Panels (d), (e), and (f) illustrate the logarithmic disparity between the ground state energies determined by both QAOA approaches and those ascertained through precise diagonalization of the Hamiltonian, corresponding to the three protein-ligand docking cases. For each case, we generated and optimized 500 unique sets of initial parameters to demonstrate variance, denoted by a shaded region with 50\% opacity.}
  \label{fig:QAOA_result}
\end{figure*}

\begin{figure*}
  \centering
  \includegraphics[width=1\textwidth]{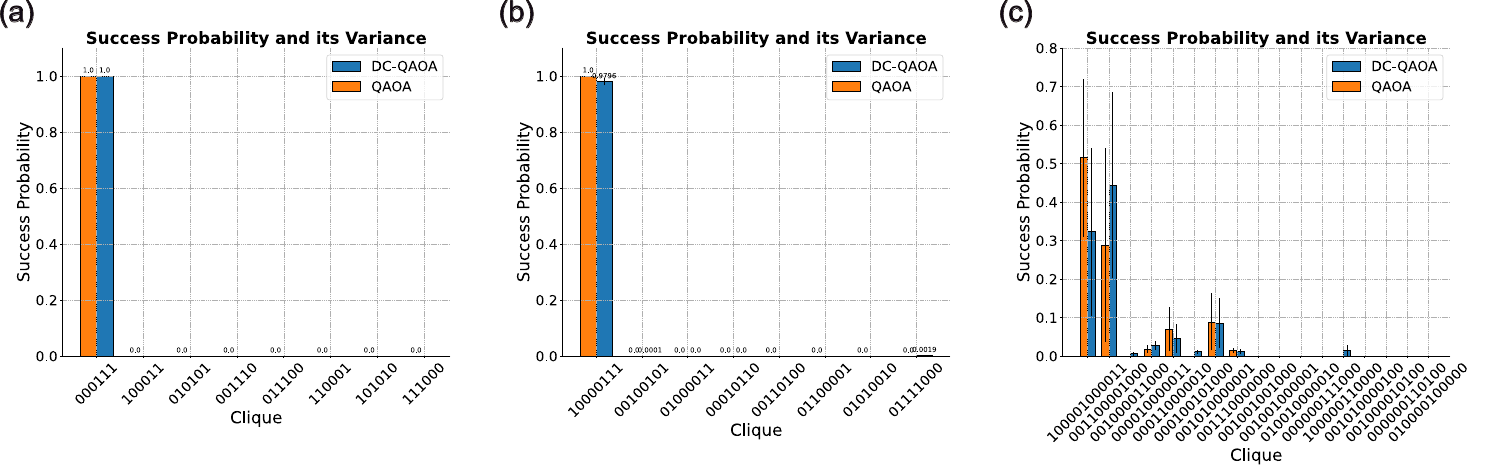}
  \caption{The success probability of docking for complex with PDB IDs: 8SKH (6 qubits), 3HAC (8 qubits), and 5F4L (12 qubits). For each case, we generated and optimized 500 unique sets of initial parameters to demonstrate variance, denoted by vertical line.}
  \label{fig:SP}
\end{figure*}

In this section, we investigate the utilization of the QAOA-based molecular-docking algorithm in expanded quantum systems of 12 qubits. 

We focus on the complex of HIV-1 gp120 and JP-III-048 (PDB ID: 5F4L), a key molecular interaction in HIV research with potential therapeutic significance. The glycoprotein HIV-1 gp120, a critical part of the human immunodeficiency virus type 1 envelope, is essential in attaching to the CD4 receptor of host cells, aiding in viral entry. JP-III-048, a compact synthetic molecule, intervenes in this process by adhering to gp120, consequently impeding the virus's capacity to infect cells and proliferate.

In this study, we have strategically identified three pharmacophore points on JP-III-048, denoted as N5302 (hd1), C5281 (hp2), and N5297 (hd3), and four distinct points on HIV-1 gp120, namely N2206 (HD1), 02209 (HA2), C5301 (HP3), and N2479 (HA4). Subsequently, we have meticulously constructed two Ligand Descriptors (LDGs), one for the JP-III-048 ligand and another for the HIV-1 gp120 receptor. The relative distances between these pharmacophores have been vividly illustrated in Figure~\ref{fig:3HAC_heat}. Furthermore, we have embarked upon the development of the Bipartite Interaction Graph (BIG), the detailed construction of which can be found in Appendix D, with its corresponding adjacency matrix depicted in Figure~\ref{fig:5f4l_interaction}(a).

To commence our analysis, we have initially applied a classical algorithm to the BIG framework, aiming to identify and rank cliques consisting of three or more vertices based on their respective weights. The graphical representation of this data is presented in Figure~\ref{fig:5F4L_cliques}. It is worth noting that a remarkable observation emerged from this analysis; the weight differential between the largest and second-largest clique was found to be quite marginal, approximately at an order of magnitude of merely $10^{-2}$. This subtle variation presents a noteworthy challenge to the effectiveness of our numerical algorithm.

Drawing upon insights derived from prior research, we have meticulously refined key parameters within our current model. These adjustments encompassed the calibration of the penalty term, denoted as $P$, and the configuration of the optimization layer. In particular, for this experimental setup, we selected a configuration with 13 layers, while rigorously fixing the penalty term at $P = 8$. The optimization process employed the quantum natural gradient descent algorithm, and comprehensive details of the experimental circuit can be found in Appendix D.

We conducted a thorough evaluation of both the DC-QAOA and the QAOA by subjecting them to simulations across 500 distinct sets of initial parameters. The results of these simulations, specifically the trends in loss functions and success probabilities, are presented graphically in FIGs.~\ref{fig:QAOA_result}(f) and~\ref{fig:SP}(c). An in-depth analysis of the loss functions reveals a significant enhancement in the algorithm's efficiency in converging towards the ground state through the incorporation of the CD term. Quantitatively, we observed that the average loss for DC-QAOA fell within the order of $10^{-1}$, as opposed to an average loss of $10^{-{1/2}}$ for the standard QAOA. This disparity is even more conspicuous in terms of success probability: the DC-QAOA algorithm achieved a noteworthy accuracy rate of 51\%, whereas the conventional QAOA algorithm exhibited a considerably lower success rate of 32\%, indicating a greater predisposition to converge towards sub-optimal solutions. These findings provide compelling evidence that the efficacy of these algorithms is potentially influenced by the number of layers chosen and the precise implementation details of the algorithm. Nevertheless, one notable caveat of our approach is the algorithm's vulnerability to converging towards local optima within the vast parameter space. This aspect of the algorithm's performance necessitates further investigation to enhance its optimization capabilities.

\section{\label{sec:citeref} Conclusion}

This work illuminates the potential of quantum computing, specifically the conventional QAOA and its variant, the digitized-counterdiabatic QAOA, in tackling the intricate problem of molecular docking. By transforming the molecular docking problem into the maximum vertex weight clique problem, we demonstrate that quantum computing offers an innovative and efficient methodology for exploring the vast search space of protein-ligand interactions. Our investigation of various biological systems, including the SARS-CoV-2 Mpro complex with PM-2-020B, the DPP-4 complex with piperidine fused imidazopyridine 34, and the HIV-1 gp120 complex with JP-III-048, underscores the adaptability of this approach. The results obtained through QAOA and DC-QAOA are in close alignment with biological principles, suggesting that quantum computing can provide invaluable insights into the design and optimization of therapeutic drugs.

Our research presents a transformative paradigm in constructing protein-ligand interaction graphs, emphasizing the pivotal role of pharmacophore interactions. This methodology has yielded consistently more accurate docking results, in alignment with established biological principles. In the realm of algorithmic development, our findings robustly demonstrate that the DC-QAOA outperforms the conventional QAOA in identifying the ground states of problem Hamiltonians. This superiority is evident in two key aspects: DC-QAOA exhibits higher accuracy at equivalent circuit depths and requires fewer iterations under the same parameter settings. The enhancement is attributed to the incorporation of the CD term in DC-QAOA, which effectively minimizes transitions to excited states, thus enhancing the precision of ground state identification. Moreover, in the current era of NISQ technologies, characterized by prevalent quantum noise, the lower quantum gate count of DC-QAOA is particularly significant. This reduced gate count in DC-QAOA facilitates the application of existing quantum devices to larger and more complex molecular systems, a crucial advancement given the challenges posed by quantum noise. 

% {}

%For instance, consider the Quantinuum H2 quantum processor, a testament to Honeywell's technological innovation, where the fidelity of single-qubit gates has reached an impressive 99.997\%, and that of two-qubit gates, 99.98\%~\cite{quantinuum_h2}. Our numerical simulations affirm the capability of quantum systems of this caliber to perform molecular docking procedures with enhanced precision in an 8-qubit quantum system.

Notwithstanding the promising outcomes achieved, our investigation into a 12-qubit system with the QAOA-based algorithm has encountered substantial challenges. These impediments are likely attributed to the exponential increase in optimization parameters, a complexity that potentially impairs the efficacy of our optimization algorithm. Notably, even the integration of Counterdiabatic (CD) terms does not entirely mitigate this issue. The intrinsic optimization challenge within QAOA remains a pivotal area for further exploration, as novel optimization strategies could be instrumental in transcending local optima.

In recent years, the field of quantum computing has experienced significant advancements, heralding a potential shift beyond the Noisy Intermediate-Scale Quantum (NISQ) era. Remarkable progress has been made in a variety of quantum systems, including superconducting~\cite{cao2023generation}, trapped-ion~\cite{quantinuum_h2}, neutral atomic~\cite{bluvstein2023logical}, and optical quantum systems~\cite{madsen2022quantum}. These advancements are of paramount importance as they facilitate the manipulation of an increased number of qubits with enhanced precision, laying the groundwork for revolutionary breakthroughs across multiple scientific fields. For example, the Quantinuum H2 quantum processor, a testament to Honeywell's technological ingenuity, has achieved single-qubit gate fidelity of 99.997\% and two-qubit gate fidelity of 99.98\%~\cite{quantinuum_h2}. Our numerical simulations substantiate the potential of quantum systems of this caliber to execute molecular docking procedures with superior precision, even within an 8-qubit quantum framework.

%it is crucial to recognize that 

In our proposed quantum algorithm for the docking problem,  we have incorporated multiple approximation methodologies to confront this multifaceted challenge. As advancements in quantum technology continue, a diminished reliance on these approximation methods is anticipated. This progression is poised to yield improvements in both the precision and efficiency of quantum algorithms, particularly as we approach the post-NISQ era. This forthcoming era undeniably demands sustained and pioneering advancements in the field of quantum algorithms and their accompanying software frameworks. This scenario underscores the critical need for interdisciplinary collaboration and ongoing research endeavors. The fusion of diverse expertise, combined with cutting-edge innovations, represents a pivotal element in fully harnessing the potential of quantum computing. Such integration of resources and knowledge is positioned at the vanguard of reshaping the paradigms within various scientific and industrial domains, ultimately exerting a profound influence on the future trajectory of these disciplines.

In conclusion, our study emphasizes the transformative potential of quantum computing in the realm of molecular docking, thereby enhancing our comprehension of molecular interactions. This progress has significant ramifications for the development of effective therapeutic strategies. We argue that incorporating quantum computing throughout the drug discovery process presents a significant opportunity to initiate a paradigm shift in biomedical research.

%These advancements are not confined to the domains of chemical and material science simulations; they extend to optimization in logistics and finance, novel approaches in machine learning, and breakthroughs in cryptography. Overcoming challenges related to qubit error rates and scalability is essential to realize these potential advancements. 

% {As we transition from the NISQ era, quantum computing is poised to gain significant advantages in several domains. Key areas of potential quantum advantage include chemical and material science simulations, optimization problems in logistics and finance, advancements in machine learning, and cryptography. Overcoming challenges related to qubit error rates and scalability remains crucial. The evolution beyond NISQ will necessitate ongoing innovations in quantum algorithms and software, and a synergistic approach between quantum and classical computing. This transition marks a significant step in the maturation of quantum technology, highlighting the need for interdisciplinary collaboration and continuous research.}

\begin{acknowledgments}
%Mark Fingerhuth and Leonardo Banchi for their valuable discussions and suggestions. We acknowledge
We would like to express our gratitude to Dr. Shang Yu for providing supplementary materials for their related paper Ref.~\cite{yu2022universal}. This work is supported by the National Natural Science Foundation of China Grant (No.~12175003 and No.~12361161602),  NSAF (Grant No.~U2330201). The scientific calculations in this paper have been done on the HPC Cloud Platform of Shandong University, and supported by the High-performance Computing Platform of Peking University.

Q.D. and Y.H. contributed equally to this work.

\end{acknowledgments}

\bibliography{apssamp}% Produces the bibliography via BibTeX.

\newpage
\appendix
\begin{widetext}
\section{Quantum natural gradient descent algorithm}
As the natural gradient adjusts the learning rates for different parameters based on the curvature of the cost function, it enables more efficient and faster convergence during training. The idea behind the natural gradient algorithm is to compute gradients by mapping the parameter to a Riemannian manifold on the parameter space instead of Euclidean space, where each point corresponds to a parameter vector. In that space, a metric matrix can be defined, reflecting the geometric structure of the parameter space. Unlike gradient descent algorithms, the natural gradient algorithm differs by scaling the gradient with Fisher information matrix, instead of $\mathbb{I}$, as the metric matrix $\bm{M}$. Thereby, it provids better control over the updating direction of model parameters.
For efficiently training the model, we utilize the quantum natural gradient descent as the optimizer which leverages the quantum Fisher information matrix to adjust learning rates for variational parameters in quantum circuits\cite{stokes2020quantum}. The parameter update rule of $\bm{\theta}$ in $i$-th iteration shows as follows, 
\begin{equation}
    \bm{\theta}^{(i+1)} = \bm{\theta}^{(i)} - \eta\cdot \bm{M^{-1}C},
\end{equation}
where $\bm{C}\in \mathbb{C}^n, \bm{M}\in \mathbb{C}^{n\times n}$
\begin{align}
    &\bm{C}_j = \partial_{\theta_j}\langle H \rangle_{\bm{\theta}}, \\
    &\bm{M}_{jk} = \mathcal{R}[\langle \partial_{\theta_j}\psi|\partial_{\theta_k}\psi\rangle - \langle \partial_{\theta_j}\psi|\psi\rangle\langle\psi|\partial_{\theta_k}\psi\rangle].
\end{align}
Here, unlike the matrix $\bm{M}$ is associated with the Fisher information matrix in natural gradient descent, it is related to the quantum geometric tensor (QGT) which is the Fubini-Study metric on quantum states in such quantum counterpart.
We perform all the numerical experiments of DC-QAOA for molecular docking under $\eta=0.1$ and randomly sample the gate parameters from a uniform distribution, $\bm{\theta}\sim U(0,1)$ to initialize the circuit.

\section{Supplementary Materials--8SKH}

In this investigation, we enhance the $\tau$ flexible approach introduced in \cite{banchi2020molecular} by adjusting the parameter $\tau + 2\epsilon$ to $\tau + \epsilon_1 + \epsilon_2$. This adaptation permits the customization of specific interaction distances tailored to distinct pharmacophore categories. Precisely, for vertices composed exclusively of hydrogen bond acceptors and donors, we employ a shorter interaction distance, denoted as $\epsilon_{s}$ \AA. Conversely, for all other scenarios, we maintain the standard interaction distance of $\epsilon$ \AA.

The determination of the connectivity within the BIG is calculated utilizing the subsequent equation:

\begin{equation}
In_{strA-strB} = (\tau + \epsilon_A + \epsilon_B) - |d_{1} - d_{2}|.
\end{equation}

In this study, the variable \( In_{strA-strB} \) denotes the connectivity strength of a given edge within the BIG. This term is determined based on the ratio of \( strA \) and \( strB \), which represent the names of the two vertices forming an edge. Additionally, \( d_1 \) and \( d_2 \) are defined as the Euclidean distances of the ligand and the protein, respectively, located at the vertices of the edge in question.

Our proposed methodology involves a matrix representation, wherein each element corresponding to a pair of vertices is assigned a value based on the connectivity strength. Specifically, an element is set to 1 if \( In_{strA-strB} > 0 \), indicating the presence of an edge between the vertices. Conversely, an element is assigned a value of 0 if \( In_{strA-strB} \leq 0 \), signifying the absence of a connection. It is crucial to note that all diagonal elements of this matrix are set to 0. This specification is deliberately made to preclude the counting of self-interactions, thereby maintaining the integrity and accuracy of the interaction analysis in the BIG framework.

In 6-qubits example, considering the elements ha1-HD1, ha1-HD2, ha1-HD3, hp2-HD1, hp2-HD2, and hp2-HD3, and selecting values of $\epsilon = 4 $ \AA, $\epsilon_s = 3 $ \AA, and $\tau = 0.6$ \AA, we can calculate the values for BIG. These values are visually represented in Fig.~\ref{fig:8SKH_heat}, and the corresponding adjacency matrix is illustrated in Fig.~\ref{fig:8skh_interaction}(c).

\begin{figure}[htbp]
  \centering
  \includegraphics[width=0.95\textwidth]{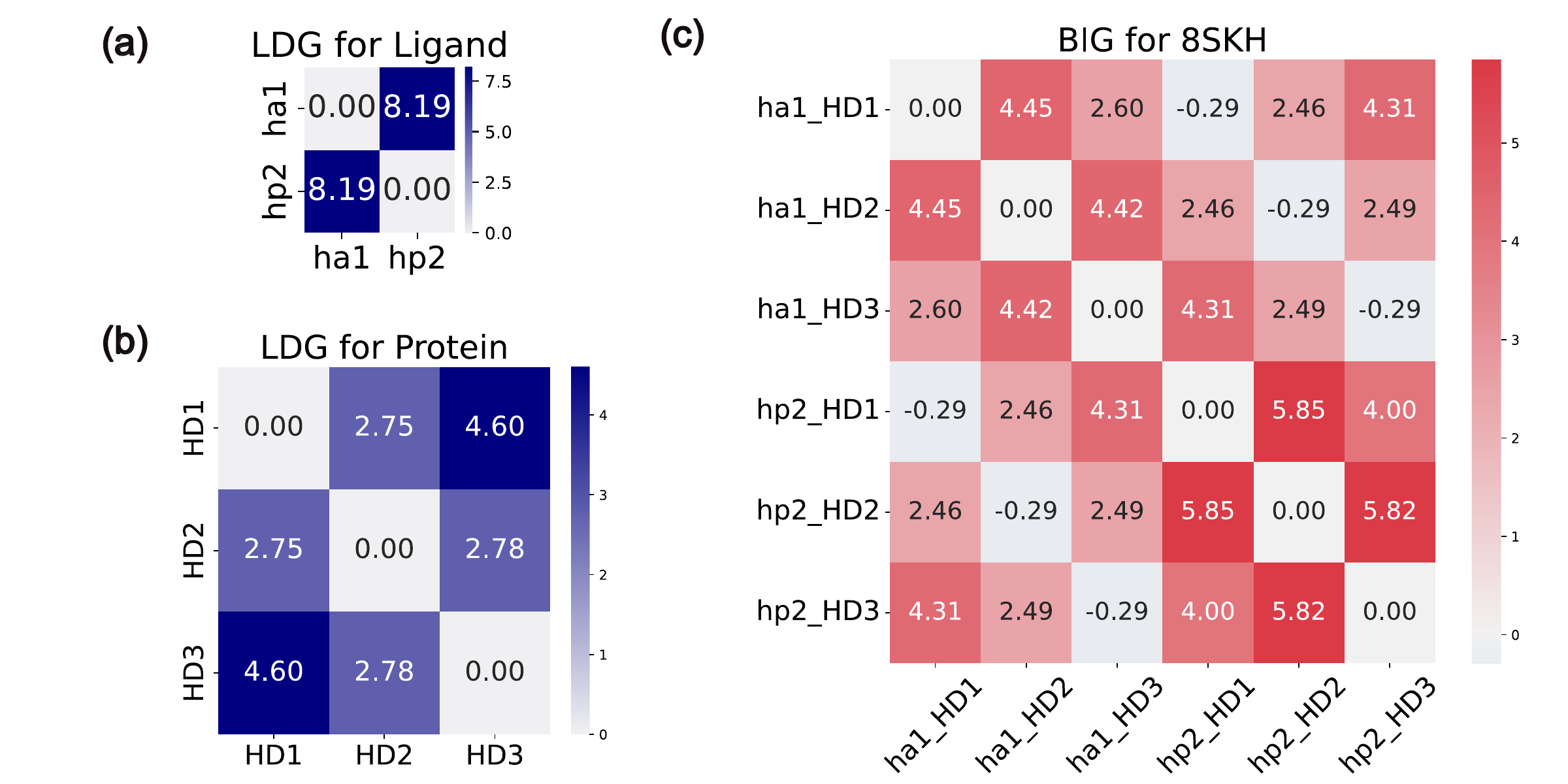}
  \caption{Panel (a) and Panel (b) present the distance data for the protein and ligand, respectively. Panel (c) illustrates the original data for the BIG of 8SKH, highlighting two pharmacophore points on PM-2-020B (O2378 (ha1) and C2373 (hp1)) and three pharmacophore points on SARS-CoV-2 Mpro (N1104 (HD1), N1108 (HD2), and N1114 (HD3)).}
  \label{fig:8SKH_heat}
\end{figure}

\begin{figure}[htbp]
  \centering
  \includegraphics[width=0.9\textwidth]{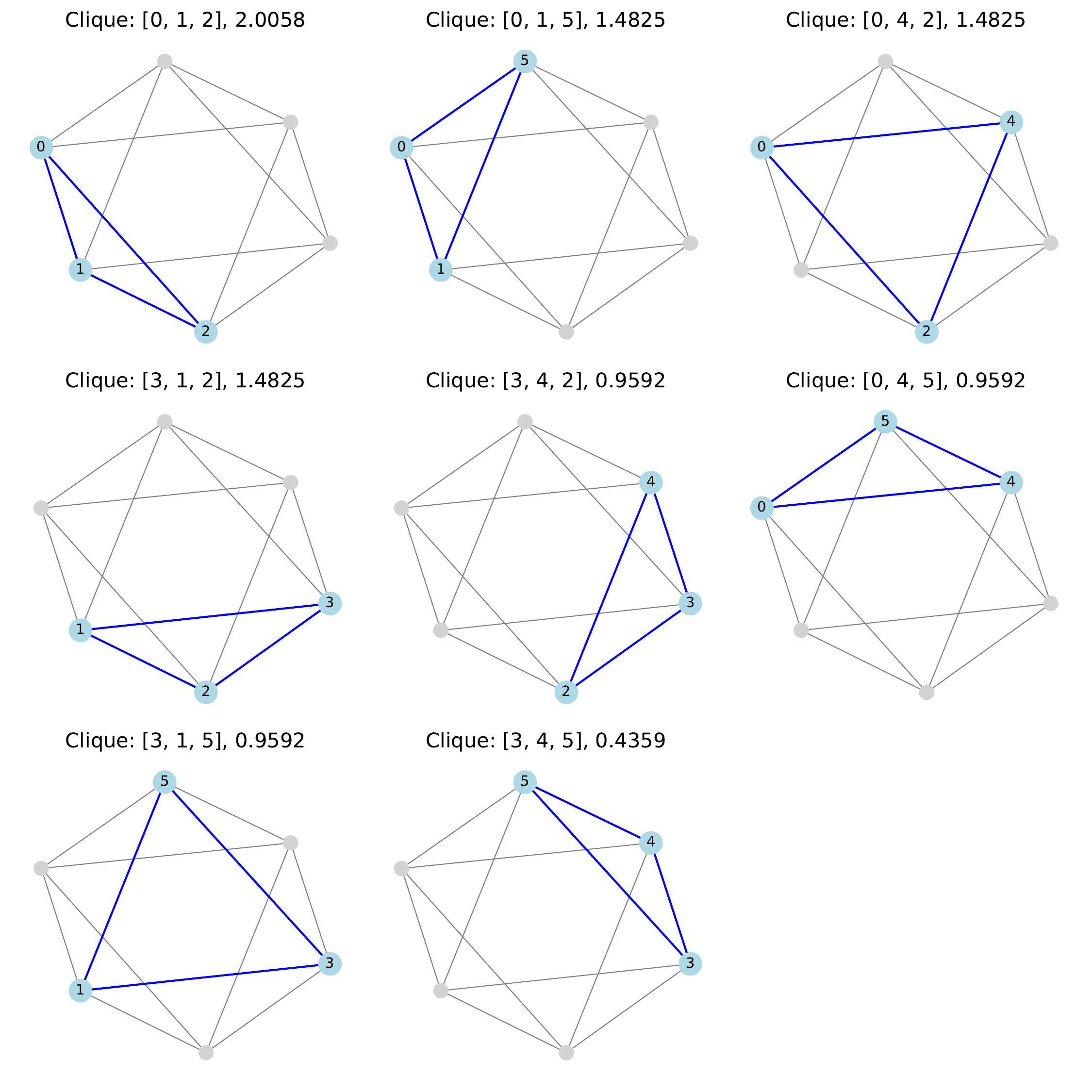}
  \caption{All cliques with a greater number of vertices than 3 in BIG for 8SKH}
  \label{fig:8SKH_cliques}
\end{figure}

Furthermore, we present the quantum circuits of DC-QAOA and QAOA, as depicted in Fig.~\ref{fig:8SKH_circuit}. It is noteworthy that the QAOA circuit does not incorporate the $U_{CD}$ component. The "number of layers" in the QAOA algorithm context refers to the frequency of repetition of the sequence highlighted in the yellow box within the circuit diagram.

\begin{figure}[H]
  \centering
  \includegraphics[width=0.8\textwidth]{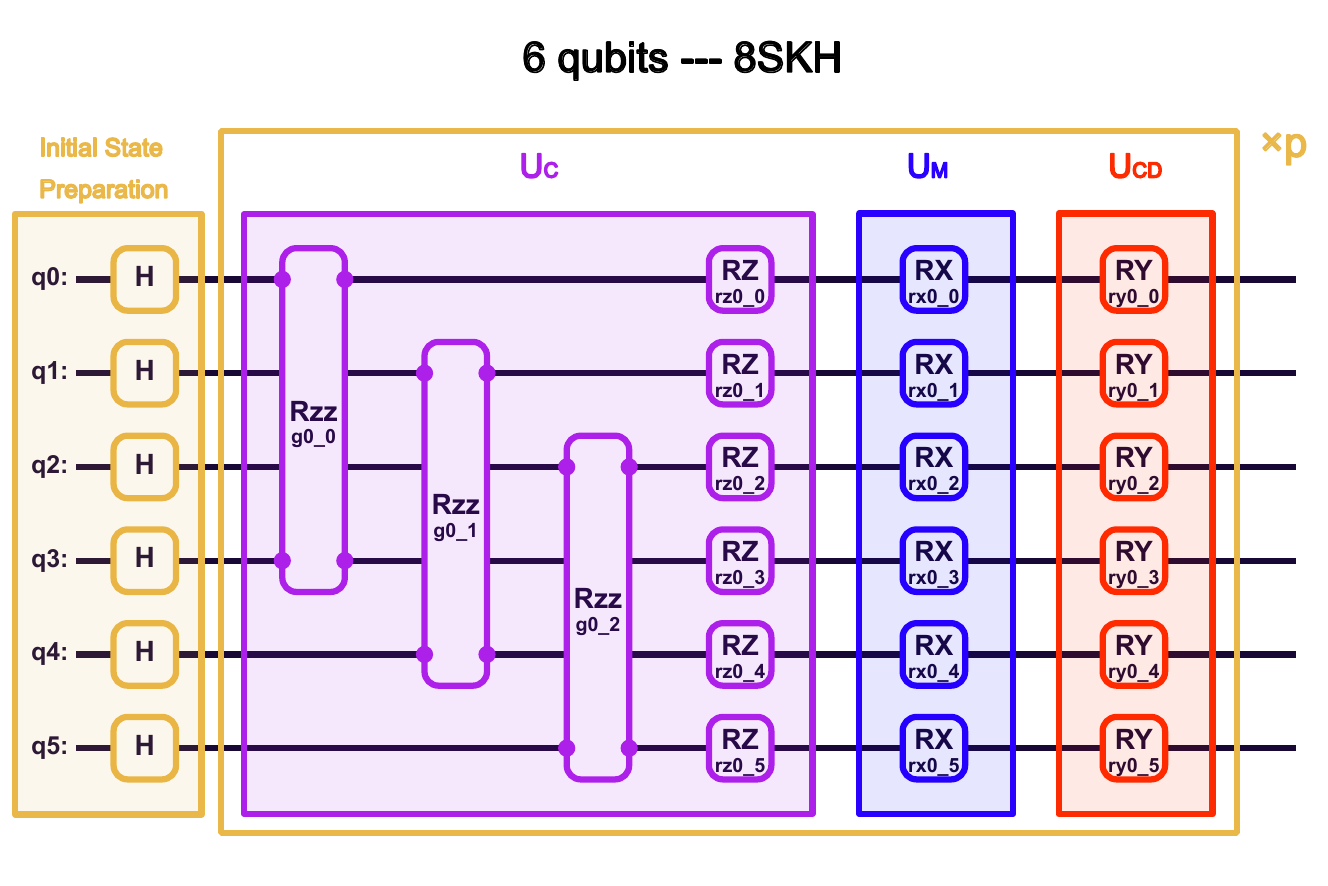}
  \caption{The quantum circuit configuration for the QAOA-based algorithm for 8SKH. The quantum circuit is partitioned into distinct segments, each responsible for a separate phase of the algorithm's execution. Initial quantum states are engineered into coherent superpositions through the application of the Hadamard gate (H) to each qubit.Following the state preparation, the circuit enters the control unitary (\( U_C \)) phase, characterized by the application of entangling two-qubit ZZ rotations (Rzz) and single-qubit Z rotations (RZ). Subsequent to \( U_C \), the mixing unitary (\( U_M \)) is implemented using single-qubit X rotations (RX) for each qubit. The circuit culminates with the controlled-drive unitary (\( U_{CD} \)), which consists of single-qubit Y rotations (RY), whereas QAOA does not have $U_{CD}$ part. The entire sequence encapsulates 'p' iterations.}
  \label{fig:8SKH_circuit}
\end{figure}

\section{Supplementary Materials--3HAC}

\begin{figure}[ht]
  \centering
  \includegraphics[width=0.95\textwidth]{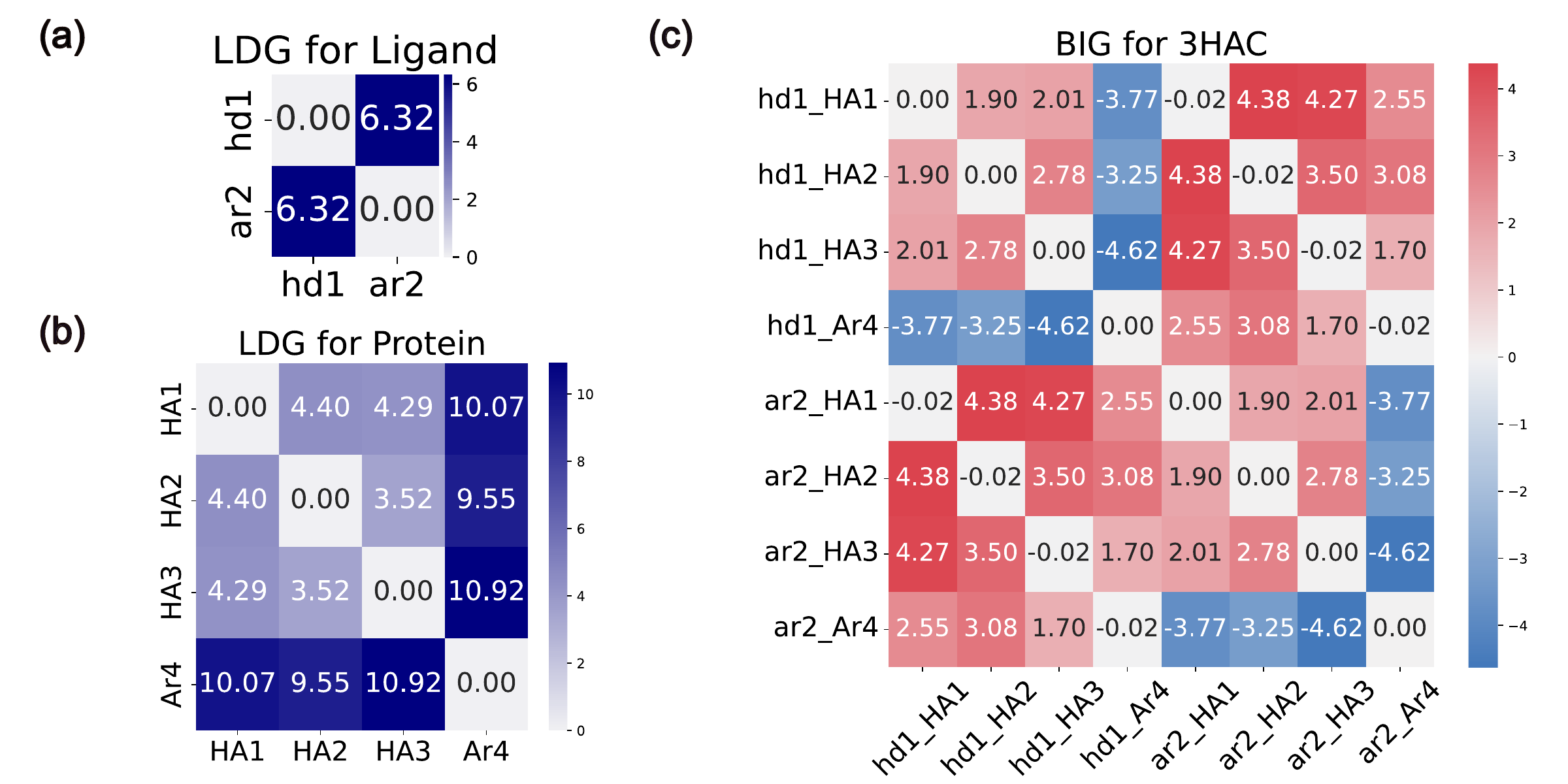}
  \caption{Panel (a) and Panel (b) present the distance data for the protein and ligand, respectively. Panel (c) illustrates the original data for the BIG of 8SKH, highlighting two pharmacophore points on the piperidine fused imidazopyridine 34 ligand: N12298 (hd1), and the pseudo-atoms at the center of the aromatic ring (ar2, the central coordinates of six atoms: C12301, C12302, C12306, N12307, C12308, and C12309). We also selected four pharmacophore points in DPP-4: 01453 (HA1), O1462 (HA2), C5171 (HA3), and the pseudo-atoms in the center of the aromatic ring (AR4, the central coordinates of six atoms: C2683, C2684, C2685, C2686, C2687, and C2688)}
  \label{fig:3HAC_heat}
\end{figure}

In the investigation of the 8-qubit system, our analysis concentrated on the interactions involving hd1-HA1, hd1-HA2, hd1-HA3, hd1-Ar4, ar2-HA1, ar2-HA2, ar2-HA3, and ar2-Ar4. For this specific configuration, the parameters were meticulously defined as $\epsilon = 3.1$ \AA, $\epsilon_s = 2.5$ \AA, and $\tau = 0.1$ \AA. This approach represents a deviation from the conventional 6-qubit model. The primary rationale for this deviation stems from the spatial compactness inherent in the selected pharmacophore points, which necessitated a precise adjustment of the parameters to accommodate the denser configuration. Following these parameter modifications, the BIG values were computed. The results of these computations are graphically represented in Fig.\ref{fig:3HAC_heat}. In addition, the corresponding adjacency matrix, which provides an insightful visualization of the intermolecular interactions, is effectively depicted in Fig.\ref{fig:3hac_interaction}(a).

\begin{figure}[ht]
  \centering
  \includegraphics[width=0.95\textwidth]{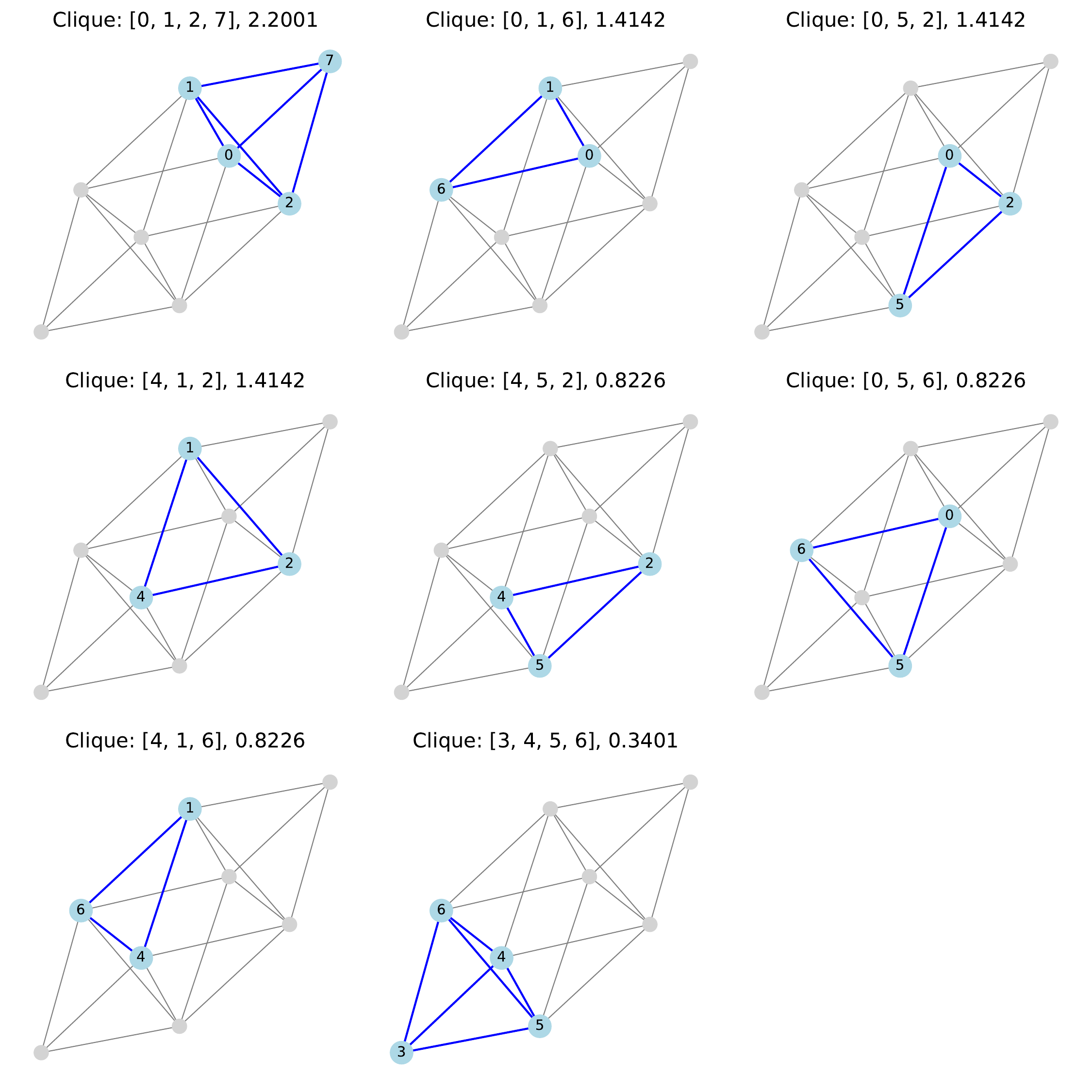}
  \caption{Illustration of all cliques exceeding three vertices in the BIG for 3HAC.}
  \label{fig:3HAC_cliques}
\end{figure}

\begin{figure}[ht]
  \centering
  \includegraphics[width=0.95\textwidth]{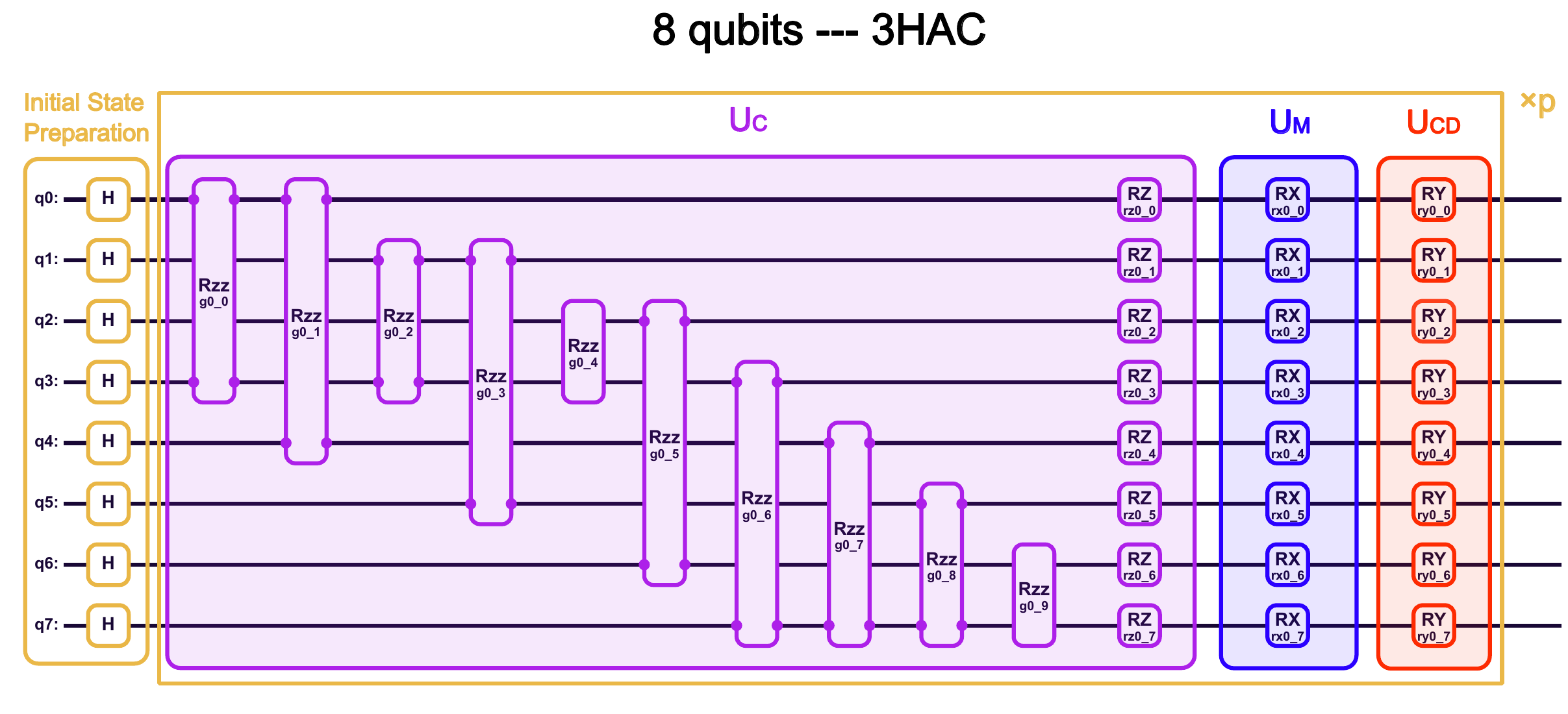}
  \caption{Quantum Circuit Configuration for the QAOA-based algorithms applied to 3HAC. This circuit is segmented into distinct phases, each integral to the algorithm's operation. Initial quantum states are configured into coherent superpositions via the Hadamard gate (H) application on each qubit. Subsequently, the circuit progresses to the control unitary (\( U_C \)) phase, marked by entangling two-qubit ZZ rotations (Rzz) and single-qubit Z rotations (RZ). Following \( U_C \), the mixing unitary (\( U_M \)) is executed, employing single-qubit X rotations (RX) for each qubit. The circuit concludes with the controlled-drive unitary (\( U_{CD} \)), comprising single-qubit Y rotations (RY), a component not present in the standard QAOA. This entire process encompasses 'p' iterations.}
  \label{fig:3HAC_circuit}
\end{figure}

\begin{figure}[ht]
  \centering
  \includegraphics[width=0.95\textwidth]{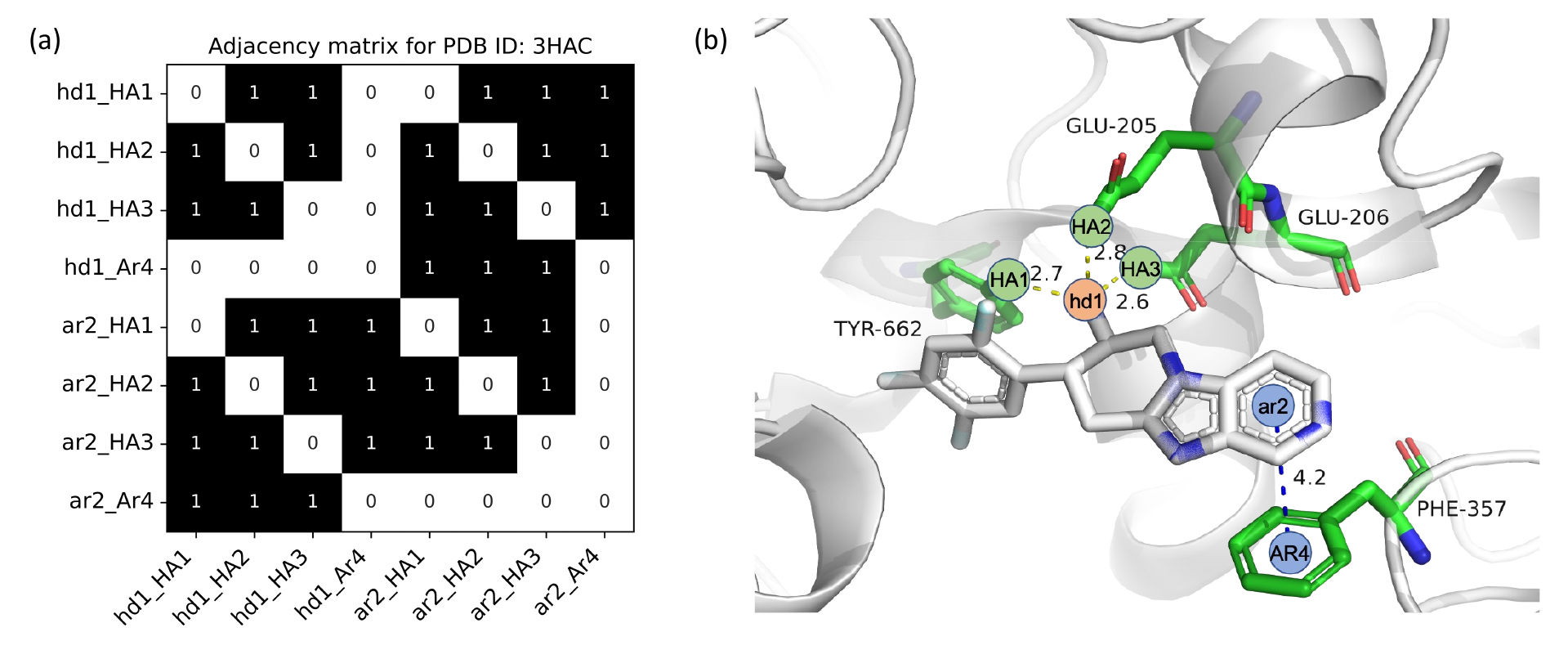}
  \caption{Panel~(a) depicts the adjacency matrix of the binding interaction graph for DPP-4 in complex with piperidine fused imidazopyridine 34 (PDB ID: 3HAC). Panel~(b) presents the Predicted Protein-Ligand Docking Visualization, showcasing interacting amino acid residues on the protein (green carbon atoms) and the ligand (white carbon atoms). Notably, the yellow bonds highlight crucial predicted interactions, namely hd1- HA1, hd1-HA1, hd1-HA1, and ar2-AR4.}
  \label{fig:3hac_interaction}
\end{figure}

Furthermore, this study delineates the quantum circuits employed for both the DC-QAOA and traditional QAOA algorithms, as exemplified in Fig.~\ref{fig:3HAC_circuit}. A pivotal distinction in the QAOA circuit configuration is the omission of the $U_{CD}$ component. Within the QAOA algorithmic framework, the "number of layers" is defined as the count of sequential repetitions of the process enclosed in the yellow box in the circuit diagram.

\section{Supplementary Materials-5F4L}

\begin{figure}[ht]
  \centering
  \includegraphics[width=0.95\textwidth]{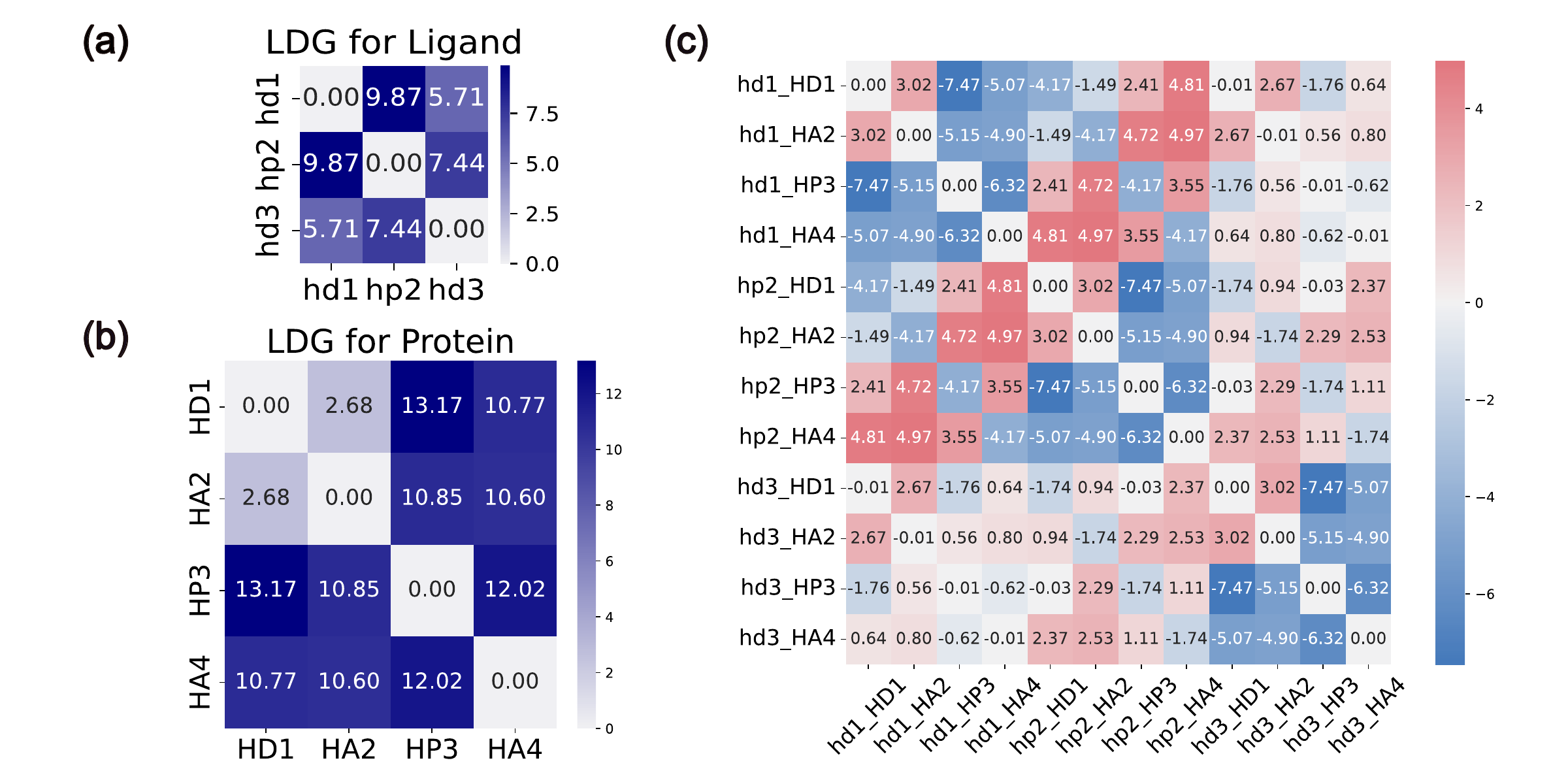}
  \caption{Panel (a) and Panel (b) present the distance data for the protein and ligand, respectively. Panel (c) illustrates the original data for the BIG of 8SKH, highlighting three pharmacophore points on JP-III-048: N5302 (hd1), C5281 (hp2), and N5297 (hd3), and four points on HIV-1 gp120: N2206 (HD1), 02209 (HA2), C5301 (HP3), and N2479 (HA4). }
  \label{fig:5F4L_heat}
\end{figure}

In the context of a 12-qubit system, we designate the parameters $\epsilon = 2.8$ \AA, $\epsilon_s = 2.5$ \AA, and $\tau = 0.1$ \AA. These values facilitate the computation of the Binding Interaction Graph (BIG), as delineated in Fig.~\ref{fig:5F4L_heat}. Correspondingly, the resulting adjacency matrix is elucidated in Fig.~\ref{fig:5f4l_interaction}(a).

\begin{figure}[ht]
  \centering
  \includegraphics[width=0.95\textwidth]{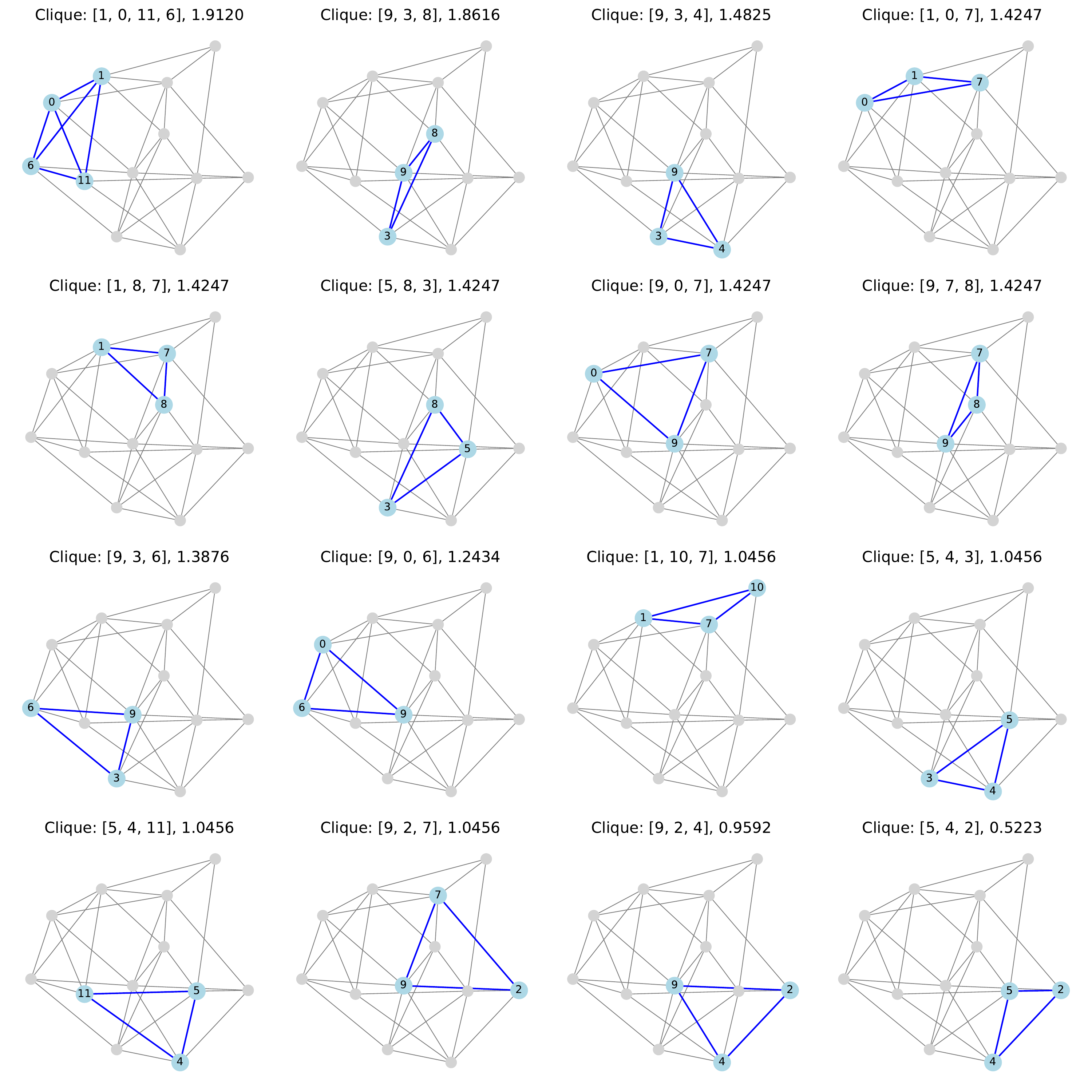}
  \caption{Visualization of all cliques in BIG for 5F4L, encompassing vertices exceeding a count of three.}
  \label{fig:5F4L_cliques}
\end{figure}

Moreover, we expound upon the quantum circuit architectures of both the DC-QAOA and the standard QAOA, as depicted in Fig.~\ref{fig:5F4L_circuit}. It is imperative to note that the QAOA circuit configuration excludes the inclusion of the $U_{CD}$ component. Within the algorithmic framework of QAOA, the 'number of layers' is indicative of the iteration frequency of the sequence encapsulated within the circuit diagram's highlighted yellow box.

\begin{figure}[ht]
  \centering
  \includegraphics[width=0.8\textwidth]{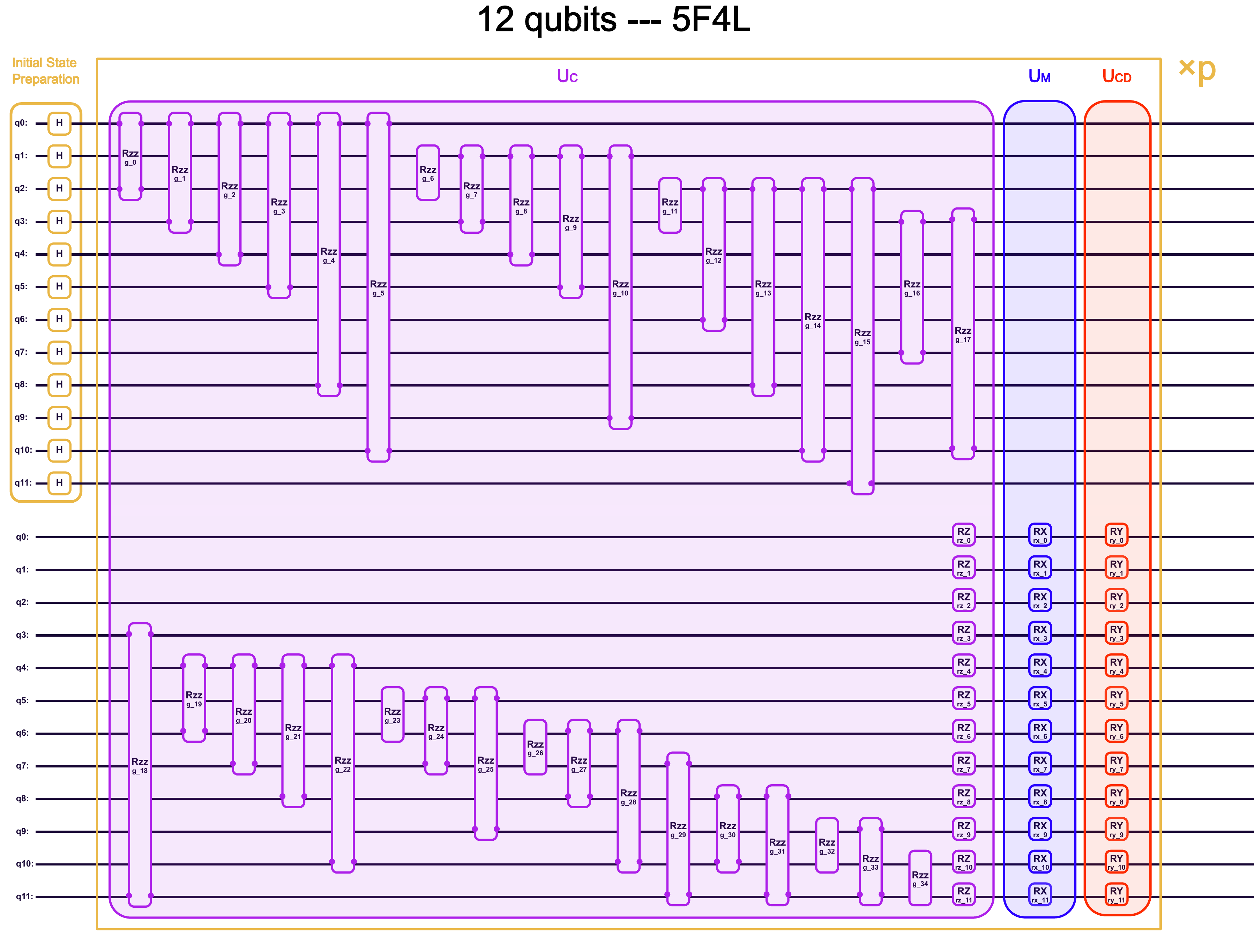}
  \caption{DC-QAOA quantum circuit configuration for 5F4L, segmented into distinct operational phases. The circuit initiates with the generation of initial quantum states, transforming each qubit into a coherent superposition via the Hadamard gate (H). The ensuing phase involves the control unitary ($U_C$), characterized by a series of entangling two-qubit ZZ rotations (Rzz) and single-qubit Z rotations (RZ). Following $U_C$, the mixing unitary ($U_M$) is executed, employing single-qubit X rotations (RX) for each qubit. The circuit culmination is marked by the implementation of the controlled-drive unitary ($U_{CD}$), constituted by single-qubit Y rotations (RY), a segment absent in the standard QAOA circuit. This sequence encompasses a total of 'p' iterations.}
  \label{fig:5F4L_circuit}
\end{figure}

\begin{figure}
  \centering
  \includegraphics[width=0.8\textwidth]{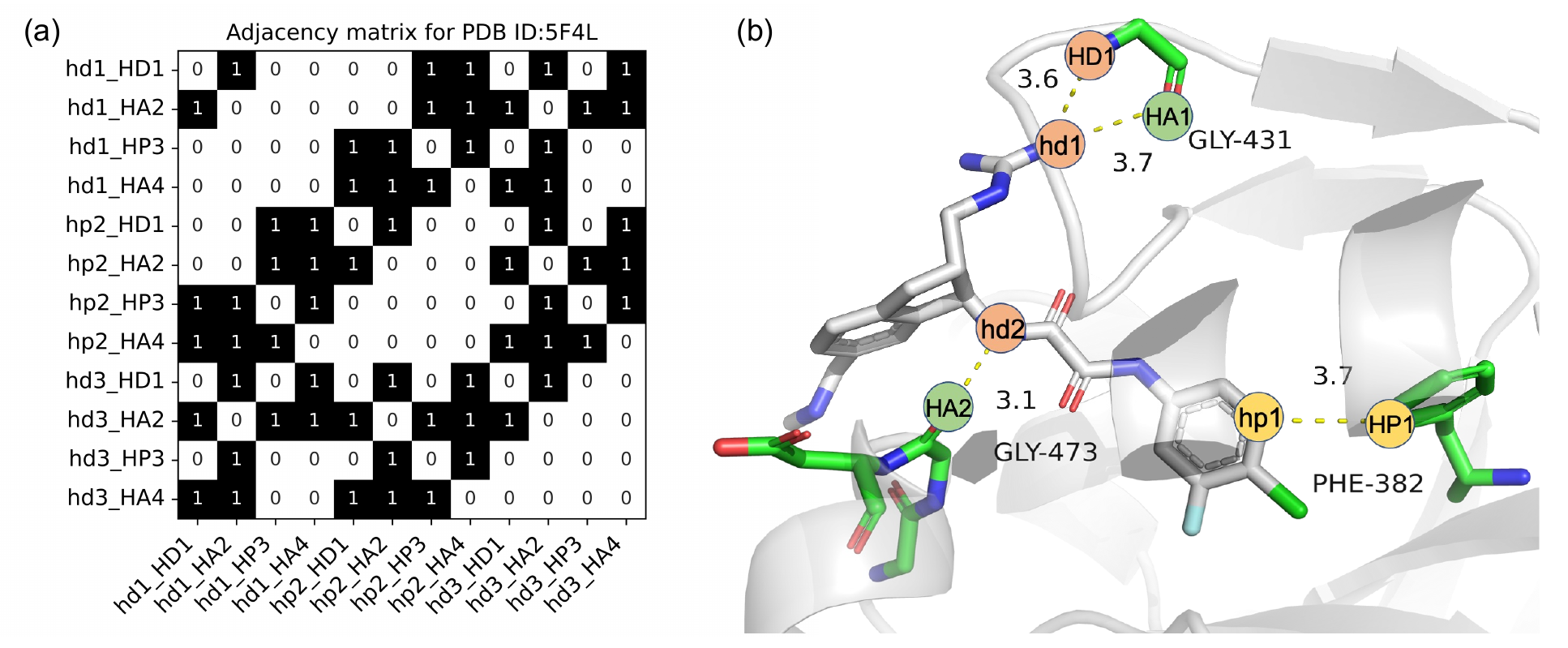}
  \caption{(a) The adjacency matrix of the binding interaction graph of the HIV-1 gp120 complex with JP-III-048. (b) Predicted Protein-Ligand Docking Visualization: the interacting amino acid residues on the protein are shown with green carbon (C) atoms, while the ligand carbon atoms are depicted in white. The highlighted yellow bonds signify the predicted interactions, specifically, hd1-HD1, hd1-HA1, hd2-HA2, and HP-HP1. Notably, hd1 functions as both a hydrogen bond donor and acceptor, connecting to HD1 and HA1 due to the positive charge on its nitrogen atom [Ng+]. Under these circumstances, the nitrogen atom primarily serves as a hydrogen bond donor, while its capacity to act as a hydrogen bond acceptor might be diminished.}
  \label{fig:5f4l_interaction}
\end{figure}
\end{widetext}

%\renewcommand\thefigure{A\arabic{figure}}

%\begin{figure}[htbp] 
%\centering 
%\includegraphics[width=0.48\textwidth]{Picture1.png} 
%\caption{(a) 2D Structure of JP-III-048: Comprehensive visualization of pharmacophore features, comprising five hydrogen bond donors (orange), three hydrogen bond acceptors (green), five hydrophobic groups (yellow), and two aromatic rings (blue). (b) 3D Structure of JP-III-048: Depiction of selected pharmacophore elements engaged in molecular docking, along with the pairwise Euclidean distances (unit: angstrom).} 
%\label{Figmain2}
%\end{figure}

\end{document}